\documentclass[twocolumn,twoside]{IEEEtran}

\usepackage{mathpple}
\usepackage{times}

\usepackage{amsmath}  % Define \boldsymbol (in amsbsy too) and align

\usepackage{amssymb}  % Define \mathbb (in amsfonts too)
\usepackage{mathrsfs} % Define \mathscr, a script font

\usepackage{theorem}  % Helps in rendering theorems, etc.
\usepackage{cite}     % Gives multiple references as intervals
\usepackage{comment}  % Comments out with \begin{comment} ... \end{comment}
\usepackage{hyperref}

\usepackage{upref}
\usepackage{amsfonts}

\usepackage{verbatim}

\usepackage{color}
\usepackage[usenames,dvipsnames]{xcolor}

\usepackage{graphicx}
\usepackage{tikz}
\usepackage{nccmath}

\usepackage{latexsym}
\usepackage[ruled,vlined,linesnumbered]{algorithm2e}
\usepackage[draft,ulem=normalem]{changes}
\definechangesauthor[color=red]{fs}
\setremarkmarkup{\footnote{#1: \textcolor{Changes@Color#1}{#2}}}
\usepackage{makecell}

\usepackage{enumitem}

\allowdisplaybreaks

\newcommand{\be}[1]{\begin{equation}\label{#1}}
\newcommand{\ee}{\end{equation}}

\newcommand{\bc}{\begin{center}}
\newcommand{\ec}{\end{center}}

\newcommand{\floor}[1]{\lfloor{#1}\rfloor}

\newcommand{\cC}{{\cal C}}

\newcommand{\cN}{{\cal N}}

\renewcommand{\leq}{\leqslant}

\renewcommand{\geq}{\geqslant}

\makeatletter
\newcommand{\pushright}[1]{\ifmeasuring@#1\else\omit\hfill$\displaystyle#1$\qquad\fi\ignorespaces}
\newcommand{\pushleft}[1]{\ifmeasuring@#1\else\omit$\displaystyle#1$\hfill\fi\ignorespaces}
\makeatother

\newcommand{\Cref}[1]{Co\-rol\-la\-ry\,\ref{#1}}

\theoremstyle{plain} \theorembodyfont{\normalfont\slshape}

\newtheorem{thm}{Theorem$\!$}
\newenvironment{theorem}{\begin{thm}\hspace*{-1ex}{\bf.}}{\end{thm}}

\newtheorem{prop}[thm]{Proposition$\!$}

\newtheorem{lem}[thm]{Lemma$\!$}
\newenvironment{lemma}{\begin{lem}\hspace*{-1ex}{\bf.}}{\end{lem}}

\newtheorem{cor}[thm]{Corollary$\!$}

\newtheorem{defi}[thm]{Definition$\!$}

\newtheorem{cl}[thm]{Claim$\!$}
\newenvironment{claim}{\begin{cl}\hspace*{-1ex}{\bf .}}{\end{cl}}

\theorembodyfont{\normalfont}

\newtheorem{exam}{Example$\!$}
\newenvironment{example}{\begin{exam}\hspace*{-1ex}{\bf .}}{\end{exam}}

\newtheorem{remrk}{Remark$\!$}

\newtheorem{const}{Construction$\!$}

\newcommand*{\medcup}{\mathbin{\scalebox{1.5}{\ensuremath{\cup}}}}%

\begin{document}

% paper title
\title{Exact Reconstruction from Insertions in Synchronization Codes}

%\author{\large Frederic~Sala,~\IEEEmembership{Student Member,~IEEE,} Clayton~Schoeny,~\IEEEmembership{Student Member,~IEEE,} Ryan~Gabrys,~\IEEEmembership{Member,~IEEE,} and Lara Dolecek,~\IEEEmembership{Senior Member,~IEEE}

\author{\large Frederic~Sala,~\IEEEmembership{Student Member,~IEEE,} Ryan~Gabrys,~\IEEEmembership{Member,~IEEE,} Clayton~Schoeny,~\IEEEmembership{Student Member,~IEEE,} and Lara Dolecek, \IEEEmembership{Senior Member,~IEEE}

\thanks{F. Sala, C.~Schoeny, and L. Dolecek are with the Electrical Engineering Department, University of California, Los Angeles, Los Angeles, CA 90095, USA. (e-mail: \texttt{ \{fredsala, cschoeny\}@ucla.edu, dolecek@ee.ucla.edu}).}
\thanks{R. Gabrys is with Spawar Systems Center Pacific Code 532, San Diego, CA, 92152 (e-mail: \texttt{ryan.gabrys@navy.mil}).}
\thanks{Research supported in part by the NSF Graduate Research Fellowship Program and NSF grant CCF-1527130. Part of the results in this paper were presented at the IEEE International Symposium on Information Theory (ISIT) in 2015 and 2016 (references~\cite{SalaISIT,SalaISIT2}).}
\thanks{Copyright (c) 2016 IEEE. Personal use of this material is permitted.  However, permission to use this material for any other purposes must be obtained from the IEEE by sending a request to pubs-permissions@ieee.org.}}
\maketitle

\bibliographystyle{IEEEtranS}

\begin{abstract}
This work studies problems in data reconstruction, an important area with numerous applications. In particular, we examine the reconstruction of binary and nonbinary sequences from synchronization (insertion/deletion-correcting) codes. These sequences have been corrupted by a fixed number of symbol insertions (larger than the minimum edit distance of the code), yielding a number of distinct traces to be used for reconstruction. We wish to know the minimum number of traces needed for exact reconstruction. This is a general version of a problem tackled by Levenshtein for uncoded sequences.

We introduce an exact formula for the maximum number of common supersequences shared by sequences at a certain edit distance, yielding an upper bound on the number of distinct traces necessary to guarantee exact reconstruction. Without specific knowledge of the codewords, this upper bound is tight. We apply our results to the famous single deletion/insertion-correcting Varshamov-Tenengolts (VT) codes and show that a significant number of VT codeword pairs achieve the worst-case number of outputs needed for exact reconstruction. We also consider extensions to other channels, such as adversarial deletion and insertion/deletion channels and probabilistic channels.

\end{abstract}

\section{Introduction}
This paper is concerned with the problem of reconstructing sequences (selected from error-correcting codes) from {\color{black} many corrupted copies, referred to as} traces. {\color{black}In particular, we are concerned with sequences from insertion/deletion-correcting codes that have been corrupted by insertion errors.} {\color{black} We are interested in answering the following question. If we are given a codeword from an {\color{black}$(\ell-1)$}-insertion/deletion-correcting $q$-ary code of length $n$, what is the minimum number $M$ of distinct traces (produced by a $t$-insertion channel) that always allows for exact reconstruction of the codeword?}

Our main result is that exact reconstruction is always possible when $M$ is at least %M$ distinct tracesof a length-$n$ $q$-ary codeword in an {\color{black}$(\ell-1)$}-insertion/deletion-correcting code for %$M$ at least
{ %\small
\begin{equation}  \label{eq:formula}
 \sum_{j={\color{black}\ell}}^t \sum_{i=0}^{t-j} \binom{2j}{j} \binom{t+j-i}{2j} \binom{n+t}{i}(q-1)^i (-1)^{t+j-i}+1.
\end{equation}}
Without further knowledge of the code, the bound in \eqref{eq:formula} is tight. In other words, if $M$ is smaller than the right hand side of \eqref{eq:formula}, there exists a pair of sequences of length $n$ with $M$ common traces, so that reconstruction cannot be guaranteed, and, moreover, these sequences are at edit distance at least $2\ell+1$ (we make this terminology precise later on), and could thus be part of an {\color{black}$(\ell-1)$}-insertion/deletion-correcting code. The result generalizes the uncoded version from \cite{Lev4} and \cite{Lev3}, which can be recovered by taking $\ell=1$ in \eqref{eq:formula} and simplifying the result through a series of combinatorial identities. It is surprising that exact formulas like \eqref{eq:formula} exist, given the paucity of exact expressions in insertion and deletion problems. Before we further discuss results such as \eqref{eq:formula}, we briefly introduce the context for this work.

%  It also generalizes the authors' partial binary result in \cite{SalaISIT}

Data reconstruction from traces is an important problem with numerous applications, including data recovery, genomics and other areas of biology, chemistry, sensor networks, and many others. The general problem of reconstruction is broadly divided into probabilistic and adversarial variants. In the probabilistic version, the traces are formed by passing the data through a noisy channel (typically an edit channel with certain deletion and insertion probabilities) and the goal is to reconstruct the data to within a certain error probability. In\cite{Batu}, an algorithm is introduced based on bitwise majority alignment that reconstructs an original sequence of length $n$ with high probability from $O(\log n)$ traces when the deletion probability is $O(\log \frac{1}{n})$. These results were extended to the deletion/insertion channel in \cite{Kannan} and improved upon in \cite{Viswanathan}. Sequence reconstruction with constant deletion probability was also studied in \cite{Holenstein}, where the authors showed that when the sequence length is $n$, reconstruction is possible with high probability from a number of traces polynomial in $n$ in time polynomial in $n$.

The adversarial variant, which we are concerned with in this paper, allows for traces formed from a worst-case number of errors and seeks to determine what is the smallest number of traces needed for zero-error reconstruction \cite{Lev4}. This setup for sequence reconstruction has also been applied to associative memories \cite{Yaakobi2012}. In these memories, each entry is associated with neighboring entries; when searching for a particular entry, ``clues'' are given in the form of such neighboring entries. This notion leads to a generalization of sequence reconstruction; here, the question becomes how many sequences are associated with (i.e., of maximum Hamming distance from) three or more sequences. The resulting intersection is called an \emph{output set}, and the size of the maximum output set is the \emph{uncertainty} of the memory. This line of research was extended in \cite{Junnila14}, which studies efficient codes for information retrieval in memories with small uncertainty, and in \cite{Junnila16}, where the number of input clues is varied. We note that all of these works target the Hamming metric. 
%Junnila and Laihonen
%The probabilistic version of the reconstruction problem for the deletion channel with fixed probability was studied
%Applications to biology are particularly popular; recent works in this vein include \cite{Acharya, Kiah} and \cite{Shomorony}.

By contrast, we are specifically interested in the following problem: if a codeword from a synchronization (specifically, insertion/deletion-correcting) code, i.e., a code with a certain minimum edit distance, is repeatedly transmitted through a noisy channel, how many distinct channel outputs (traces) are necessary for zero-error reconstruction? This question is indeed meaningful; consider, for example, phylogenomics, where we wish to reconstruct the genetic sequence of an ancestor organism from a large number of sequences of evolutionary descendants. Each of the descendant sequences is formed from a number of base pair insertions. The related question of how to efficiently perform the reconstruction is tackled from a coding-theoretic point of view in the recent work \cite{Farzad}.

%from shorter strings (traces) produced by sequencing devices. For example, \cite{Acharya} explores a similar topic in the context of protein sequencing. A critical question is how many traces are needed to ensure an accurate reconstruction.

Moreover, reconstruction of encoded data has a natural application to data storage. We can partition the operational lifetime of a disk drive, memory, or other data storage device into two periods. The first period is regular, short-term use; popular devices and common error-correcting codes all target this scenario. Here, a small number (often one) of channel outputs are used to read the data. The second period refers to extremely long-term use of the device, well beyond the guaranteed operating lifetime. In this case, many reads can be performed, resulting in a large number of traces that can be used to recover the original data. This type of very long term use is increasingly relevant. For example, DNA storage is targeted as a medium to store data for $10^4$ or more years, and, indeed, over the long term, DNA sequences are affected by insertions and other errors that can be modeled by insertions (duplications, tandem/block duplications, block insertions) \cite{Benson}. The present paper studying reconstruction from insertions can therefore be viewed as complementary to the many recent efforts studying coding for data storage in DNA \cite{Jain, Moshe2, RyanDameraru, Yazdi, RyanLee, Kiah, Yazdi2, Grass, Bornholt}. 

{\color{black} Our work also joins recent coding-theoretic studies on insertions and deletions, such as \cite{Liron, Kulkarni, Cullina2}. Our main result \eqref{eq:formula}, as we shall see, computes a maximal common intersection of error balls centered at codewords. A similar result, but in the easier case of codes correcting substitution errors, has been used to provide improved Gilbert-Varshamov bounds in \cite{Jiang} and \cite{Vu}, in the cases of binary and non-binary codes, respectively. Our work can thus be seen as a building block towards providing improve cardinality bounds for insertion/deletion-correcting codes. }

%As an example, we can reconstruct the complete genomes of extinct organisms from millions of years ago. It is thus not surprising that DNA storage is targeted as a medium to store data for $10^4$ or more years \cite{XXX}.

In a classic paper, \cite{Lev4}, Levenshtein explored several variations of the reconstruction problem, studying both adversarial and probabilistic channels and exact and approximate reconstruction. However, the problem of reconstructing sequences affected by insertions and deletions in the case where the sequences are themselves part of a code (e.g., have a certain minimum edit distance) was left open. We tackle this problem for the insertion case in the current work. We target insertions for two reasons. First, insertions (and deletions) are edit errors, which are of particular interest as we often wish to reconstruct strings. In keeping with our biology theme, as described above, we note that insertions are a common type of mutation affecting genetic sequences. Second, unlike deletions, insertions offer symmetries that allow a tractable search for exact formulas. 

%The rest of this paper is organized as follows. First, we discuss prior work, our problem setup, and notation in Section II. We include a proof of results such as the one in \eqref{eq:formula} in Section III. We apply the result to the single insertion/deletion-correcting Varshamov Tenengolts (VT) codes in Section IV.

The remainder of this paper is organized in the following way. In the next section, we introduce our problem setup, discuss prior work, and describe notation. In Section III, we prove and interpret the general, non-binary common supersequences problem. We also discuss important special cases and corollaries. In Section IV, we apply our result to the single deletion/insertion-correcting Varshamov-Tenengolts (VT) codes. Finally, in Section V, we consider extensions to the adversarial deletion and insertion/deletion channels {\color{black} and briefly consider probabilistic channels}. We conclude the paper in Section VI. 
 %In Section III, we prove a result on the common supersequences problem for the binary case. In Section IV, we prove and interpret the more general, non-binary result. We also discuss important special cases and corollaries. In Section V, we apply our result to the single deletion/insertion-correcting Varshamov-Tenengolts (VT) codes. Finally, in Section VI, we consider extensions to the deletion and insertion/deletion channels. We conclude the paper in Section VII. 

\section{Preliminaries}

\subsection{Problem setup}
Levenshtein {\color{black} provided a general framework for reconstruction problems in \cite{Lev4}. He considered the problem of reconstructing a sequence $X \in V \subseteq \mathbb{F}_q^n$ for a set $V$ and an alphabet $\mathbb{F}_q$. In this case, the traces were selected from $B_t(X,H)$, the ball produced by applying up to $t$ single errors from a set of error functions $H$ to $X$. Levenshtein noted that it is always possible to exactly reconstruct $X$ given $N^H_q(V,t)+1$ distinct elements of $B_t(X,H)$}, where $N^H_q(V,t)$ is defined by
\begin{equation}
\label{definter}
N^H_q(V,t) := \max_{X,Z \in V, X\neq Z} |B_t(X,H) \cap B_t(Z,H)|.
\end{equation}

{\color{black} The idea here is that $X$ cannot share any more than $N^H_q(V,t)$ traces with any other element of $V$, so that $N^H_q(V,t)+1$ distinct traces uniquely determine $X$. Thus}, the problem of exact reconstruction of sequences can be identified with the combinatorial problem of counting (the maximum number of) common distorted sequences. In \cite{Lev4}, expressions were given for $N^H_q(V,t)$ for many choices of error sets $H$. In \cite{Lev3}, the focus was specifically on deletion and insertion channels. In this paper, we focus on the case of insertion channels, so that $H$ is made up of single symbol insertions, and we wish to reconstruct $X$ from its supersequences (sequences formed from $X$ by insertions). In \cite{Lev4}, an expression was provided for $N^H_q(V,t)$ in this scenario, but only for the specific case of $V=\mathbb{F}_q^n$, the \emph{uncoded} case. For this problem setup, we may write the balls $B_t(X,H)$ as insertion balls $I_t(X)$ and denote the expression $N^H_q(\mathbb{F}_q^n,t)$ as $N^+_q(n,t)$. In \cite{Lev4}, $N^+_q(n,t)$ was shown to be
\begin{equation} \label{eq:Lev1}
N^+_q(n,t) = \sum_{i=0}^{t-1} \binom{n+t}{i} (q-1)^i (1-(-1)^{t-i}).
\end{equation}

However, the problem of reconstruction given a code $V$ differing from the entire set $\mathbb{F}_q^n$ was left open. We tackle this problem in the present work. Consider, for example, reconstructing a sequence that is a member of an $(\ell-1)$-insertion-correcting code $\cC'$. Sequences that are part of such codes must have a minimum edit distance of $2\ell$. If this is the case, we can always perform exact reconstruction if we know $N_q^+(\cC',t) + 1$ distinct supersequences of $X$, where

\[N_q^+(\cC',t) = \max_{X,Z \in \cC', X\neq Z} |I_t(X) \cap I_t(Z)|.\]

Computing this maximal intersection requires knowing the structure of the code $\cC'$. This is challenging, since few such codes are known outside of the single insertion case. Instead, we focus on deriving an expression for the maximum number of common supersequences for sequences at a minimum particular edit distance $2\ell$, 

\begin{figure}
\centering
\includegraphics[width=\linewidth]{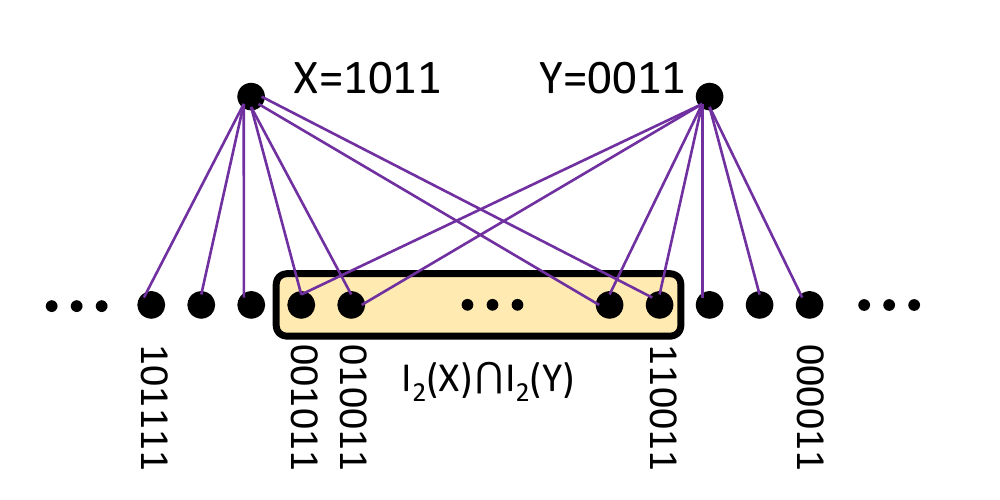}
\caption{Example setup for our problem. Here, we are counting common supersequences of $X=1011$ and $Y=0011$ produced by $t=2$ bit insertions. We have that $|I_2(X) \cap I_2(Y)| = 12$.} \label{fig:exampleFig}
\end{figure}

\[N_q^+(n,t,\ell) = \max_{\substack{X,Z \in \mathbb{F}_q^n\\ d_e(X,Z) \geq 2\ell}} |I_t(X) \cap I_t(Z)|.\]
An example is shown in Figure~\ref{fig:exampleFig} for $n=4$ and $t=2$. Specifically, we prove that $N^+_q(n,t,\ell)$ is given by
{ 
\[ \sum_{j=\ell}^t \sum_{i=0}^{t-j} \binom{2j}{j} \binom{t+j-i}{2j} \binom{n+t}{i}(q-1)^i (-1)^{t+j-i}.\]
}
Evaluating the above expression provides an upper bound on the number of channel outputs needed for reconstruction of codewords in insertion-correcting codes. Without specifying a particular code construction, this bound is the best possible. Note that Levenshtein's formula $N^+_q(n,t)$ in \eqref{eq:Lev1} can be written as $N^+_q(n,t) = \max_{\ell \geq 1} N_q^+(n,t,\ell).$
%for $\ell, \ell+1, \ldots$ yields the maximum number of common supersequences for sequences at distance $2\ell$ or greater and thus

We will see, in fact, that this maximum is always attained at $\ell = 1$. In other words, the maximum number of common supersequences occurs for sequences that are as ``close'' as possible. 

As part of our study, we provide an even more general version of this result where we allow the sequences $X$ and $Z$ to have different lengths. This result can be interpreted as a {\color{black}generalization} of Levenshtein's formula $N^+_q(n,t)$. %Special cases of this very general expression turn out to have very interesting forms; we spend some time examining these special cases and their implications.  	

%The remainder of the paper is organized as follows. In the next section, we set up notation. In Section III, we introduce our main result on the exact solution to the common supersequences problem. In Section IV, we consider an application of the result to the single insertion correcting Varshamov Tenengolts codes. We conclude the paper in Section V. 

\subsection{Notation}

%%% SECTION: BINARY RESULT
%\section{Binary Result}
%In this section, we introduce the binary common supersequences result. First, we introduce some notation and the problem setup.

%\subsection{Notation and Problem Setup}
We introduce some useful notation. Let $\mathbb{F}_q$ denote the set $\{0,1,\ldots, q-1\}$ for $q \geq 2$ and $[a,b]$ denote the set $\{a, a+1, a+2, \ldots, b-1, b\}$ if $a \leq b$. We denote sequences with upper-case letters and individual symbols with lower-case letters, so that, for example, $X = x_1x_2 \ldots x_n \in \mathbb{F}_q^n$ while $x_i \in {\color{black}\mathbb{F}_q}$ for $1\leq i \leq n$. We write $XY$ for the concatenation of sequences $X$ and $Y$; similarly, $aX$ is the concatenation of a symbol $a$ with sequence $X$. We sometimes use the notation $XS$ where $S$ is a set. In this case, $XS$ refers to the set that contains the concatenation of $X$ with all sequences in $S$, $\{XY : Y \in S\}$. 

We use the generalized binomial coefficient: for $a,b \in \mathbb{Z}$, $\binom{a}{b} = a(a-1)\ldots(a-b+1)/b!$. We assume $0! = 1$. We set $\binom{a}{b}=0$ for $b <0$. We also use the convention $\binom{a}{0} = 1$ for all $a \in\mathbb{Z}$ and $\binom{a}{b} = 0$  if $a = 0$ and $b > 0$. We sometimes rely on the useful fact that $\binom{a}{b} = 0$ if $a>0$ and $a<b$.

% , we use the following conventions: $\binom{0}{0} = 1$, $\binom{a}{b} = 0$ if $a,b \geq 0$ and $a<b$, while $\binom{a}{b} = a(a-1)\ldots(a-b+1)/b!$ for $a<0$ and $b\geq0$. Lastly, if $b < 0$, $\binom{a}{b} = 0$.

If $n \geq v {\color{black} \geq 0}$, $Z \in \mathbb{F}_q^{n-v}$ is a \emph{subsequence} of $X \in \mathbb{F}_q^n$ if $Z$ can be formed from $X$ by deleting $v$ symbols. If $n=v$, $Z$ is the empty sequence, with length 0. Similarly, for $t \geq 0$, $W \in \mathbb{F}_q^{n+t}$ is a \emph{supersequence} of $X$ if it can be formed by $t$ symbol insertions into $X$. The set of all length $n-v$ subsequences of $X$ (also called the $v$-deletion ball centered at $X$) is denoted $D_v(X)$; similarly, the set of all length $n+t$ supersequences of $X$ (the $t$-insertion ball centered at $X$) is written $I_t(X)$.

In general, the size of $D_v(X)$ depends on the sequence $X$. For example, $|D_1(X)| = \tau(X)$, where $\tau(X)$ is the number of maximal-length runs of identical symbols in $X$. On the other hand, $|I_t(X)|$ does not depend on $X$ for any $t \geq 0$. There is a formula for the size of the supersequence set that only depends on $n$,$t$, and the alphabet size $q$ \cite{Lev1},
\begin{equation}
\label{eq:insform}
|I_t(X)| := I_q(n,t) = \sum_{i=0}^t \binom{n+t}{i} (q-1)^i.
\end{equation}

The distance between sequences $X,Y$ can be measured with the \emph{edit distance} $d_e(X,Y)$. This distance is defined in the following way: $d_e(X,Y) = k$ if $k$ is the smallest number of insertions and deletions that can be used to transform $X$ to $Y$. Note that it is not necessary that $X$ and $Y$ have the same length for $d_e(X,Y)$ to be defined. For example, take $X = 00$ and $Y = 010$. Then, $d_e(X,Y) = 1$, since we require just one insertion of a 1 into $X$ to form $Y$. If $X=110$ and $Y=000$, then $d_e(X,Y)=4$. Note that our definition of edit distance does not include substitutions. Studying codes correcting insertions, deletions, and substitutions is a different, even more challenging problem.

A $t$-insertion-correcting code $\mathcal{C}$ is a subset of $\mathbb{F}_q^n$ such that if $X,Y \in \mathcal{C}$ and $X \neq Y$, then $I_t(X) \cap I_t(Y) = \emptyset$. This is equivalent to requiring that $\mathcal{C}$ has minimum edit distance\footnote{The required minimum distance is $2t+1$; however, since all the codewords in $\mathcal{C}$ are of the same length, the distance between any codeword pair must be even, since to go from one codeword to another, there must be a deletion for every insertion. Thus the minimum distance is in fact $2t+2$. For example, the minimum edit distance of the single deletion/insertion-correcting Varshamov-{\color{black}Tenengolts} codes is 4.} $2t+2$. We also note that a $t$-insertion-correcting code is also a $t$-deletion-correcting code (and also an $a$-deletion/$b$-insertion-correcting code for any pair $(a,b)$ with $a+b \leq t$). This equivalence was proved in \cite{Lev1}.

As described, we are concerned with computing the maximum number of common supersequences between sequences with edit distance of at least $2\ell$ for $\ell \geq 1$. That is, we are interested in the quantity\footnote{The ``+'' symbol denotes the fact that we are performing insertions.} $N^+_q(n,t,\ell)$ defined as 
\[N^+_q(n,t,\ell) = \max_{\substack{X,Y \in \mathbb{F}_q^n \\ d_e(X,Y) \geq 2\ell}} |I_t(X) \cap I_t(Y)|.\] 
We refer to $n$, $t$, and $\ell$ as the \emph{length}, \emph{insertion}, and \emph{distance} arguments, respectively.

Additionally, in our results, we consider a more general version of the problem where the sequences need not be of the same length. One of the two sequences ($Y$) continues to be of length $n$ while the common supersequences remain of length $n+t$. However, we allow $X$ to be of length $n+t-k$ (that is, longer than $Y$ by $t-k$ symbols). As a result, we make only $k$ insertions into $X$. Note that we now have two insertion arguments, $t$ and $k$. Similarly, the distance between $X$ and $Y$ is now required to be $t-k+2\ell$ in order to make up for the additional distance between the sequences resulting from the differing lengths. Observe that $t \geq k \geq \ell$ in this setup. The goal, then, is to compute
\[N^+_q(n,t,k,\ell) = \max_{\substack{X \in \mathbb{F}_q^{n+t-k}, Y \in \mathbb{F}_q^n \\ d_e(X,Y) \geq t-k+ 2\ell}} |I_k(X) \cap I_t(Y)|.\]
We can always retrieve $N^+_q(n,t,\ell)$ from $N^+_q(n,t,k,\ell)$ by taking $t=k$.

\subsection{Basic Claims}
We use several simple claims as building blocks in our work. First, the following fact is an immediate consequence of our definitions.

\begin{claim}
\label{cl:distance}
For $\ell \geq 1$ and $n,t,k$ non-negative integers with $t \geq k \geq \ell$, 
\[N^+_q(n,t,k,\ell) \leq N^+_q(n,t,k,\ell-1).\]
\end{claim}
\begin{IEEEproof}
Any two sequences $X,Y$ with distance at least $t-k + 2\ell$ also have distance at least $t-k+2(\ell-1)$, so therefore the maximum number of common supersequences for distance argument $\ell-1$ is at least that for distance argument $\ell$.
\end{IEEEproof}

We also have another easy fact regarding distances.
\begin{claim}
\label{cl:pmone}
Let $X' \in \mathbb{F}_q^m$ and $Y' \in \mathbb{F}_q^n$ with $m, n>0$. If $d_{e}(X',Y') = v$ and $X' = x_1X$, then \[d_{e}(X,Y') \in \{v-1,v+1\}.\]
\end{claim}
\begin{IEEEproof}
Clearly, $d_{e}(X,Y')$ cannot be smaller than $v-1$, or we could form $X$ from $Y'$ with fewer than $v-1$ operations and insert $x_1$, retrieving $X'$ in fewer than $v$ operations, a contradiction. Similarly, $d_{e}(X,Y')$ cannot be larger than $v+1$, since we can first form $X'$ from $Y'$ and then delete $x_1$ in a total of $v+1$ operations. Lastly, since $X'$ and $X$ differ in length by 1, $d_e(X',Y')$ and $d_e(X,Y')$ cannot have the same parity.
%Let us refer to the element deleted from $A$ to form $A'$ by $x$. We note that $d_{e}(A',B)  \leq \ell+1$, since we can form $B$ from $A'$ by first inserting $x$ into $A'$ and then using a total of $\ell$ deletions and insertions to take $A$ to $B$. 
%
%Similarly, $d_{e}(A',B)  \geq \ell-1$. If $d_{e}(A',B) = j < \ell-1$, we could form $B$ from $A$ by first deleting $x$ from $A$, and then using $j$ total insertions and deletions to take the resulting $A'$ to $B$, for a total of $1+j < \ell$ insertions and deletions, a contradiction. Therefore, $d_{e}(A',B) \in \{\ell -1, \ell, \ell + 1\}$.
%
%It remains to show that $d_{e}(A',B)  \neq \ell$. A simple parity argument will suffice. Let us say that the minimum number of deletions and insertions taking $A$ to $B$ is $d$ and $i$, respectively. Then, $d+i = \ell$ and $m-d+i = n$. We have that \[2i = n+\ell-m.\] Then, let us say that there exist $d'$ deletions and $i'$ insertions taking $A'$ to $B$ with $d' + i' = \ell$. Now, $A'$ has length $m-1$, so that $m-1 -d' + i' = n$, or, $2i' = n + \ell - m + 1.$ In that case, we have $2i' = 2i+1$, which is impossible. Therefore, $d_{e}(A',B)  \neq \ell$.
\end{IEEEproof}

Next, we introduce two useful results. First, we have an observation that Levenshtein originally made in \cite{Lev3}. Consider some sequence $Z=z_1z_2 \ldots z_n$. Then, $I_t(Z)$ is the union of two disjoint sets: the set of sequences starting with $z_1$ (which can be formed by placing all $t$ insertions into $z_2 \ldots z_n$) and the set of sequences starting with the element $x \in \mathbb{F}_q \setminus z_1$ (which require $x$ to be inserted at the head of $Z$, leaving $t-1$ remaining insertions into $Z$). Formally, we have that

\begin{claim}
\label{levTrickSimple}
If $Z=z_1z_2\ldots z_n \in \mathbb{F}_q^n$ is a sequence and $t \geq 1$, then,
\begin{equation} 
I_t(Z) = z_1 I_t(z_2 z_3 \ldots z_n) \medcup \cup_{x \in \mathbb{F}_q \setminus  z_1} xI_{t-1}(Z).
\end{equation}
%If $q=2$, 
%\begin{equation} 
%I_t(Z) = z_1 I_t(z_2 z_3 \ldots z_n) \medcup \bar{z_1} I_{t-1}(Z).
%\end{equation}
\end{claim}

Here, recall that $xI_{t-1}(Z)$ refers to appending all the sequences in $I_{t-1}(Z)$ to the element $x$. The idea in Claim~\ref{levTrickSimple} can be exploited to derive recursive expressions for the number of common supersequences. A variant of the following observation was used by Levenshtein in \cite{Lev3}; we provide a proof for completeness.
\begin{claim}
\label{levTrick}
Let $n$ be a positive integer, $t,k$ be non-negative integers with $t\geq k$, and $X' \in \mathbb{F}_q^{n+t-k}, Y' \in \mathbb{F}_q^n$. Write $X' = aX$ and $Y' = bY$ with $a,b \in \mathbb{F}_q$. Then, if $a=b$,
%\begin{align} \label{levTrickEqual}
%|I_k&(X') \cap I_t(Y')| = \nonumber \\
%&|I_{k}(X) \cap I_t(Y)| + (q-1) |I_{k-1}(aX) \cap I_{t-1}(aY)|.
%\end{align}
\begin{IEEEeqnarray}{rCl}
|I_k(X') &\cap& I_t(Y')| = \nonumber \\
|I_{k}&(X)& \cap I_t(Y)| + (q-1) |I_{k-1}(aX) \cap I_{t-1}(aY)|.
\label{levTrickEqual}
\end{IEEEeqnarray}
If $a\neq b$, 
\begin{align} \label{levTrickDiff}
|I_k(X') &\cap I_t(Y')| =  \nonumber\\
&|I_k(X) \cap I_{t-1}(bY)| + |I_{k-1}(aX) \cap I_t(Y)| \nonumber \\
&\quad +(q-2)|I_{k-1}(aX) \cap I_{t-1}(bY)|. 
\end{align}
%If $q=2$, the formulas \eqref{levTrickEqual} and \eqref{levTrickDiff} reduce to 
%\begin{equation}  \label{levTrickEqualBin}
%|I_k(X') \cap I_t(Y')| = |I_{k}(X) \cap I_t(Y)| + |I_{k-1}(aX) \cap I_{t-1}(aY)|,
%\end{equation}
%and
%\begin{equation} \label{levTrickDiffBin}
%|I_k(X') \cap I_t(Y')| =  |I_k(X) \cap I_{t-1}(\bar{a}Y)| + |I_{k-1}(aX) \cap I_t(Y)|,
%\end{equation} 
%respectively.
\end{claim}
\begin{IEEEproof}
First we consider the case of $a=b$. A common supersequence $W'$ of $X'=aX$ and $Y'=aY$ either starts with $a$ or an element in $\mathbb{F}_q \setminus \{a\}$. If $W'$ starts with $a$, we write $W' = aW$. Using Claim~\ref{levTrickSimple}, $W$ can be formed by $k$ insertions into $X$ and $t$ insertions into $Y$, so $W$ is a common supersequence of $X$ and $Y$. That is, it is in the set $I_k(X) \cap I_t(Y)$. Similarly, if $W' = w_1W$ where $w_1$ is one of the $q-1$ elements in $\mathbb{F}_{q} \setminus \{a\}$, it must be formed by inserting $w_1$ at the head of $X'=aX$ and at the head of $Y' = aY$. Therefore, $W$ is in ${\color{black}I_{k-1}(aX) \cap I_{t-1}(bY)}$. There are thus $(q-1)\times |I_{k-1}(aX) \cap I_{t-1}(aY)|$ choices for such supersequences $W'$. This establishes \eqref{levTrickEqual}.

For the case of $a\neq b$, if $W'=aW$, $W$ can be formed from $X'$ by inserting $k$ elements into $X$, while $W'$ can be formed from $Y'$ by inserting $a$ at the head and $t-1$ more elements into $Y'=bY$. If $W' = bW$, it is formed from $X'$ by inserting $b$ at the head and $k-1$ elements into $X'=aX$ while $W$ is formed from $Y'$ by inserting $t$ elements into $Y$. Lastly, if $W'$ starts with $w_1$, one of the $(q-2)$ elements in $\mathbb{F}_q \setminus \{a,b\}$, it is formed from $X'$ by inserting $w_1$ at the head and $k-1$ more elements into $X'=aX$ and from $Y'$ by inserting $w_1$ at the head and $t-1$ more elements into $Y'=bY$. The sets given by the three possibilities are disjoint, giving \eqref{levTrickDiff}.
\end{IEEEproof}

\section{Maximum number of common supersequences}
We are ready to prove the main result of the present work.%, the general form of Theorem~\ref{thm:bincase}. 
\begin{theorem} \label{thm:gen}
Let $n$ be a positive integer, $t, k, \ell$ be non-negative integers such that $t \geq k \geq \ell$ and $n \geq \ell$, and $q$ be an integer with $q \geq 2$. Then, 
\begingroup\makeatletter\def\f@size{9.5}\check@mathfonts
\def\maketag@@@#1{\hbox{\m@th\large\normalfont#1}}%
\begin{align} \label{maxintersectionGen}
%N^+_q(n,t,k,\ell)  = \sum_{j=\ell}^k \sum_{i=0}^{k-j} &\binom{t-k+2j}{j}\binom{t+j-i}{t-k+2j} \times \nonumber \\ 
%&\binom{n+t}{i}  (q-1)^i (-1)^{k+j-i}.
&N^+_q(n,t,k,\ell)  = \nonumber \\
&\sum_{j=\ell}^k \sum_{i=0}^{k-j} \binom{t-k+2j}{j}\binom{t+j-i}{t-k+2j} \binom{n+t}{i}  (q-1)^i (-1)^{k+j-i}.
\end{align}\endgroup
If $t=k$,
\begin{align} \label{maxintersectioneqcase}
%N^+_q(n,t,\ell) = \sum_{j=\ell}^t \sum_{i=0}^{t-j} \binom{2j}{j}& \binom{t+j-i}{2j} \times \nonumber \\
%&\binom{n+t}{i}(q-1)^i (-1)^{t+j-i}.
&N^+_q(n,t,\ell) = \nonumber\\
&\sum_{j=\ell}^t \sum_{i=0}^{t-j} \binom{2j}{j} \binom{t+j-i}{2j} \binom{n+t}{i}(q-1)^i (-1)^{t+j-i}.
\end{align}

\end{theorem}

We begin with some observations on Theorem~\ref{thm:gen}. We are more interested in the formula in \eqref{maxintersectioneqcase} compared to that in \eqref{maxintersectionGen} because most insertion/deletion-correcting codes have equal-length codewords. We will later see (from the proof of Theorem~\ref{thm:gen}) that the sequences that yield the maximum $N^+_q(n,t,\ell)$ common supersequences are those at distance precisely $2\ell$. This confirms the intuitive idea that the maximum number of common supersequences are attained by sequences as ``close'' as possible.

%It is not difficult to see that the expression in \eqref{eq:binformeq} is decreasing in $\ell$, so that the largest possible number of common supersequences is attained precisely when $\ell =1$. This is intuitive: we expect sequences that are near each other to have many more common supersequences compared to those that are far apart. 

\begin{figure}
\centering
\includegraphics[width=\linewidth]{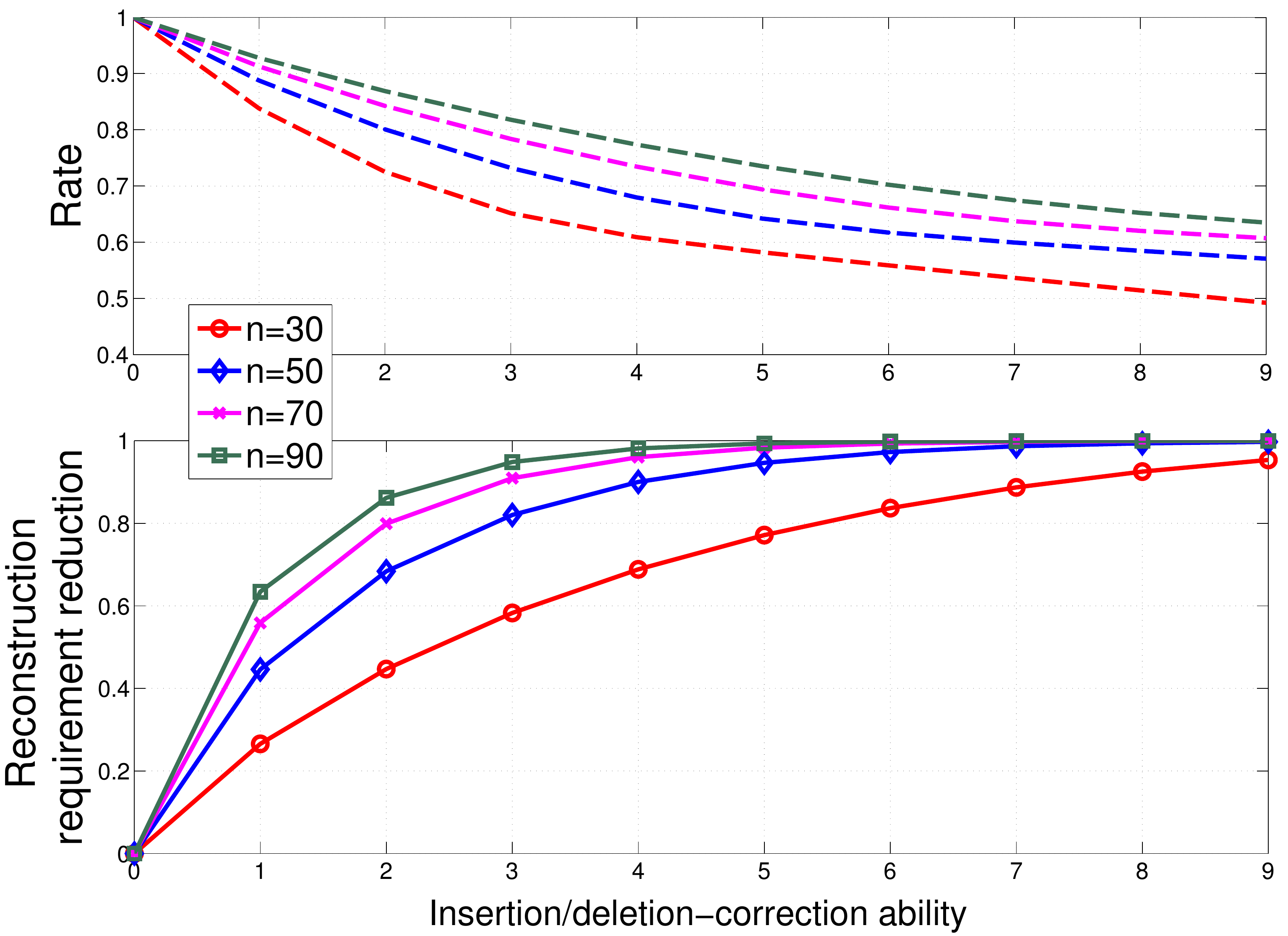}
\caption{Curves showing insertion/deletion code rates (dashed lines) and reconstruction requirement reduction percentage (solid lines) given traces affected by $t=15$ insertions for codes of lengths $n=30,50,70,90$ capable of correcting of $0,1,\ldots, 9$ edit (insertion or deletion) errors.} \label{fig:tradeoff}
\end{figure}

Results in the spirit of \eqref{maxintersectioneqcase} encourage us to examine the \textit{tradeoff between code rate and reconstruction requirements}. For example, the expression in \eqref{maxintersectioneqcase} is decreasing in $\ell$, the code's insertion/deletion-correcting ability. Increasing $\ell$ allows us to reconstruct with fewer and fewer traces, but this comes at the cost of decreased code rate. We show an example of this tradeoff for insertion/deletion-correcting codes of lengths $n=30,50,70,90$ in Figure~\ref{fig:tradeoff}. Here, we plot two curves for each code based on the error-correcting ability; the dashed curves show code rates (based on non-asymptotic upper bounds from \cite{Kulkarni}), while the thick curves show the percentage reduction (normalized to 1) in the number of traces needed to guarantee exact reconstruction given traces formed by $t=15$ random symbol insertions. %We normalize both quantities to the those of the uncoded case. %(Note that some of the terms $\binom{n+t-(2j+1)}{t-j}$ can be negative for sufficiently large $t$ and $j$, but we can only increase $\ell$ up to $n$, and in this regime, all such terms are positive.) 

%This property also yields an interesting tradeoff: a larger code minimum edit distance results in a smaller number of distinct channel outputs necessary for exact reconstruction. At the same time, a larger minimum edit distance means a worse rate for the code. In other words, we have a tradeoff between code rate and reconstruction requirements.

%Another consequence of our results is that we can recover Levenshtein's formula $N^+_2(n,t)$ in \cite{Lev3} by taking $\ell =1$ in \eqref{eq:binformeq} and applying the binomial identity \[\sum_{j=1}^t \binom{2j}{j} \binom{n+t-2j-1}{t-j} = 2\sum_{i=0}^{t-1} \binom{n+t}{i}.\]

The rest of this section is organized as follows. First, we provide the proof. Afterwards, we discuss important special cases of Theorem~\ref{thm:gen}. In particular, we recover Levenshtein's result for $N^+_q(n,t)$ from \cite{Lev4}, along with a simplified binary result as special cases. 

\subsection{Proof}
We start with a roadmap for the proof of Theorem~\ref{thm:gen}. First, we denote by $\cN^+_q(n,t,k,\ell)$ the formula 
\begingroup\makeatletter\def\f@size{9.5}\check@mathfonts
\def\maketag@@@#1{\hbox{\m@th\large\normalfont#1}}%
\begin{align*}
&\cN^+_q(n,t,k,\ell)  = \\
& \sum_{j=\ell}^k \sum_{i=0}^{k-j} \binom{t-k+2j}{j}\binom{t+j-i}{t-k+2j} \binom{n+t}{i} (q-1)^i (-1)^{k+j-i}.
\end{align*}
\endgroup
Then, our goal becomes to prove that $N^+_q(n,t,k,\ell) = \cN^+_q(n,t,k,\ell)$. We show that the formula given by $\cN^+_q(n,t,k,\ell)$ satisfies two recursions: first, $\cN^+_q(n,t,k,\ell) = \cN^+_q(n-1,t,k,\ell) + (q-1)\cN^+_q(n,t-1,k-1,\ell)$, and second, $\cN^+_q(n,t,k,\ell) = \cN^+_q(n,t-1,k,\ell) + \cN^+_q(n-1,t,k-1,\ell-1) + (q-2) \cN^+_q(n,t-1, k-1, \ell)$. This will be done purely through combinatorial manipulations of the formula given by $\cN^+_q(n,t,k,\ell)$. Afterwards, we show that the maximization given by $N^+_q(n,t,k,\ell)$ satisfies two nearly identical inequalities depending on whether the first symbol in a pair of maximizing sequences is identical or differs. We do this by exploiting Claim~\ref{levTrick}. These two results are applied in an inductive argument to show that $N^+_q(n,t,k,\ell) \leq \cN^+_q(n,t,k,\ell)$. All that remains is to give a pair of sequences that yield equality in this formula. We will show that  \[X=\underbrace{00\ldots0}_{t-k + n \text{ 0's}} \text{ and } Y = \underbrace{11\ldots 1}_{\ell \text{ 1's}}\underbrace{00 \ldots 0}_{n-\ell \text{ 0's}}.\] 
are two such sequences.

We briefly discuss two important improvements of our proof technique compared to that of Levenshtein for the $N^+_q(n,t)$ result. First, we generalize the problem to the different-lengths case where $t$ need not be equal to $k$. This enables us to involve the second recursion (for sequences starting with different symbols) directly in the induction. This was not possible in Levenshtein's proof, as the second recursion immediately breaks down into different-length cases (and formulas for such cases were not known); however, for $\ell=1$, this issue can be overcome. In the cases $\ell > 1$, this is not possible. Interestingly, our approach mirrors some proofs in combinatorics, where it is easier to prove a general formula compared to a special case. In addition, we note that unlike in Levenshtein's proof, we require a careful accounting of the recursions' effects on the distance.

%The proof of Theorem~\ref{thm:gen} uses the same approach as that of the binary version; however, the underlying recursions are more complex. We denote by 
%Thus, it is again our goal to show that $N^+_q(n,t,k,\ell) = \cN^+_q(n,t,k,\ell)$. 

%The formula $\cN^+_q(n,t,k,\ell)$ satisfies two recursions. 
\begin{lemma} \label{lem:recGen}
For $n \geq 1$, $q \geq 2$ and $t,k,\ell \geq 1$ with $t \geq k \geq \ell$, $\cN^+_q(n,t,k,\ell)$ satisfies the recursions
\begin{align*}
\cN^+_q&(n,t,k,\ell) = \\
&\cN^+_q(n-1,t,k,\ell) + (q-1)\cN^+_q(n,t-1,k-1,\ell),
\end{align*}
and
\begin{align*}
&\cN^+_q(n,t,k,\ell) = \\
&\quad \cN^+_q(n,t-1,k,\ell) + \cN^+_q(n-1,t,k-1,\ell-1) \\
&\qquad+ (q-2) \cN^+_q(n,t-1, k-1, \ell) .
\end{align*}
\end{lemma}
%Note that unlike in the binary case, the first recursion has a $(q-1)$ factor for the second term. The second recursion also contains a third term not found in the binary version. 
We defer the proof of Lemma~\ref{lem:recGen} to the appendix. {\color{black}We prove that the quantity} $N^+_q(n,t,k,\ell)$ satisfies similar recursions:

%%%%%%%%%%%%%%% LEMMA MAXIM. %%%%%%%%%%%%%%%%%%%%%%%%%%%%%%%%%%%%%%%
\begin{lemma}
\label{lem:recsmax}
Let $n$ and $q \geq 2$ be positive integers and $t, k, \ell$ be non-negative integers such that $t \geq k \geq \ell$. Let $X' =aX ,Y' = bY$ be two sequences satisfying $X' \neq Y'$, $X' \in \mathbb{F}_{q}^{n+t-k}, Y' \in \mathbb{F}_q^{n}$, and $d_{e}(X',Y') = t-k+2\ell$. Then, if $a=b$,
\begin{align*}
|I_k(X') &\cap I_t(Y')| \leq \\
&N^+_q(n-1,t,k,\ell)+(q-1)N^+_q(n,t-1,k-1,\ell),\end{align*}
and if $a \neq b$, 
\begin{align*}&|I_k(X') \cap I_t(Y')| \leq \\
&\quad N^+_q(n,t-1,k,\ell)+N^+_q(n-1,t,k-1,\ell-1) \\
&\qquad+ (q-2)N^+_q(n,t-1,k-1,\ell).
\end{align*}
\end{lemma}
\begin{IEEEproof}
We are given $X',Y'$ satisfying $X' \neq Y'$, $X' \in \mathbb{F}_{q}^{n+t-k}, Y' \in \mathbb{F}_q^{n}$, and $d_{e}(X',Y') = t-k+2\ell$. We have $X' = aX$ and $Y' = bY$, with $a,b \in \mathbb{F}_q$. In the case $a=b$, from \eqref{levTrickEqual}, 
\begin{align}  \label{eq:twoterms1}
|I_k&(aX) \cap I_t(bY)| = |I_k(aX) \cap I_t(aY)| = \nonumber \\
& |I_{k}(X) \cap I_t(Y)| + (q-1) |I_{k-1}(aX) \cap I_{t-1}(aY)|.
\end{align}

%The argument matching is identical to that in the proof of Lemma~\ref{lem:recsbinarymax}. We have that $d_{e}(X',Y') = t-k+2\ell$ and $X'=aX$, $Y'=aY$, so that $d_{e}(X,Y) =t-k+2\ell$. We have that $|I_{k}(X) \cap I_t(Y)| \leq N^+_q(n-1,t,k,\ell)$ and $|I_{k-1}(aX) \cap I_{t-1}(aY)| \leq N^+_q(n,t-1,k-1,\ell)$. Putting this into \eqref{eq:twoterms1} gives 

We note that since $d_{e}(X',Y') = t-k+2\ell$ and $X'=aX$, $Y'=aY$, $X$ and $Y$ must be at the same distance as $X'$ and $Y'$, so that $d_{e}(X,Y) = t-k+2\ell$. Thus, $X$ is of length $(n-1)+t-k$, $Y$ is of length $n-1$, and $d_{e}(X,Y) = t-k+2\ell$. Then, we have that $|I_{k}(X) \cap I_t(Y)| \leq N^+_q(n-1,t,k,\ell)$. Similarly, we have that $|I_{k-1}(aX) \cap I_{t-1}(aY)| \leq N^+_q(n,t-1,k-1,\ell)$. We call this step \emph{argument matching}, since we are computing the length, insertion, and distance arguments in order to produce the correct $N^+$ bound. Putting the two bounds into \eqref{eq:twoterms1} gives 
\begin{align*}|I_k(aX) &\cap I_t(bY)|  \leq \\
&N^+_q(n-1,t,k,\ell) + (q-1)N^+_q(n,t-1,k-1,\ell).
\end{align*}

The next case is $a \neq b$. Then, from \eqref{levTrickDiff},
\begin{align} \label{eq:twoterms2}
&|I_k(aX) \cap I_t(bY)| = \nonumber \\ 
&\quad |I_k(X) \cap I_{t-1}(bY)| + |I_{k-1}(aX) \cap I_t(Y)| \nonumber \\
&\qquad+  (q-2)|I_{k-1}(aX) \cap I_{t-1}(bY)|.
\end{align}

Again, we bound the terms in \eqref{eq:twoterms2} with formulas of the type $N^+_q(n,t,k,\ell)$. In this case, the argument matching is slightly more complicated. For the first term in \eqref{eq:twoterms2}, $bY$ has length $n$ while the insertion arguments are clearly $t-1$ and $k$. It remains to find the distance argument $\ell'$. Recall that $d_{L}(aX,bY) = t-k+2\ell$. We have, from Claim~\ref{cl:pmone}, that $d_{L}(X,bY) \in \{t-k+2\ell-1, t-k+2\ell+1\}$ and $d_{L}(aX,Y) \in \{t-k+2\ell-1, t-k+2\ell+1\}$. We must have $d_{L}(X,bY) = (t-1)-k+2\ell'$, so that $\ell' = \frac{1}{2}(d_e(X,bY) +k-(t-1)) \in \{\ell, \ell+1\}$. Thus, the possible argument tuples for the first term are $\{(n,t-1,k,\ell), (n,t-1,k,\ell+1)\}$. Next, for the second term in \eqref{eq:twoterms2}, $Y$ has length $n$ and the insertion arguments are $t$ and $k-1$. The distance argument $\ell'$ satisfies $d_{L}(aX,Y) = t-(k-1)+2\ell' \in \{t-k+2\ell-1, t-k+2\ell+1\}$ by Claim~\ref{cl:pmone}, so that $\ell' \in \{\ell-1,\ell\}$. Then, the possible argument tuples for this term are $\{(n-1,t,k-1, \ell-1), (n-1,t,k-1, \ell)\}$. Finally, the arguments for the last term must be $(n,t-1,k-1,\ell)$. Applying Claim~\ref{cl:distance}, we have that

%Since $d_{e}(aX,bY) = t-k+2\ell$, we have, from Claim~\ref{cl:pmone}, that $d_{e}(X,bY) \in \{t-k+2\ell-1, t-k+2\ell+1\}$ and $d_{e}(aX,Y) \in \{t-k+2\ell-1, t-k+2\ell+1\}$. Again, we bound the terms in \eqref{eq:twoterms2} with formulas of the type $N^+_q(n,t,k,\ell)$. The argument tuples are given by $\{(n,t-1,k,\ell), (n,t-1,k,\ell+1)\}$ for the first term, $\{(n-1,t,k-1, \ell-1), (n-1,t,k-1, \ell)\}$ for the second term, and $(n,t-1,k-1,\ell)$ for the last term. Thus,

%The two options depend on whether $d_{e}(X,bY)$ is $t-k+2\ell-1$ or $t-k+2\ell+1$. Similarly, for the second term, the parameters are in $\{(n-1,t,k-1, \ell-1), (n-1,t,k-1, \ell)\}$. Finally, the parameters are $(n,t-1,k-1,\ell)$ for the last term. We write, using Claim~\ref{cl:pmone}, % The parameter sets are given by $\{(n,t-1,k,\ell), (n,t-1,k,\ell+1)\}$ for the first term. The two options depend on whether $d_{e}(X,bY)$ is $t-k+2\ell-1$ or $t-k+2\ell+1$. Similarly, for the second term, the parameters are in $\{(n-1,t,k-1, \ell-1), (n-1,t,k-1, \ell)\}$. Then, applying Claim~\ref{cl:distance}, we have that
\begin{align*}
|I_k(X) &\cap I_{t-1}(bY)|  \\
&\leq\max \{N^+_q(n,t-1,k,\ell), N^+_q(n,t-1,k,\ell+1)\}  \\
&=N^+_q(n,t-1,k,\ell),
\end{align*}
\begin{align*}
&|I_{k-1}(aX) \cap I_t(Y)| \\ 
&\enskip \leq \max \{N^+_q(n-1,t,k-1,\ell-1), N^+_q(n-1,t,k-1,\ell) \} \\
&\enskip= N^+_q(n-1,t,k-1,\ell-1),
\end{align*} and, finally,
\begin{align*}
|I_{k-1}(aX) \cap I_{t-1}(bY)| \leq N^+_q(n,t-1,k-1,\ell).
\end{align*}

Plugging these inequalities into \eqref{eq:twoterms2} yields
\begin{align*}
&|I_k(aX) \cap I_t(bY)|  \leq \\
&\quad N^+_q(n,t-1,k,\ell) +  N^+_q(n-1,t,k-1,\ell-1)  \\
&\qquad + (q-2)N^+_q(n,t-1,k-1,\ell),
\end{align*}
and we are done.
\end{IEEEproof}

We proceed with the proof of Theorem~\ref{thm:gen}.

\begin{IEEEproof}
We first use induction on $n+t+k$ to show that
\begin{equation} \label{clmInter}
N^+_q(n,t,k,\ell) \text{ } \leq  \text{ }  \cN^+_q(n,t,k,\ell).
\end{equation}

The base cases are $n+t+k \in \{1,2\}$. First we consider $n\in \{1,2\}$ and $t=k=\ell = 0$. Since $I_0(X') \cap I_0(Y') \subseteq \{X'\}$, we have that $|I_0(X') \cap I_0(Y')| \leq 1$. The right hand side of \eqref{clmInter} evaluates to $\binom{0}{0}\binom{0}{0}\binom{n+0}{0} = 1$, as we wished. The other possibility is $n=1, t=1, k=0$, and $\ell=0$. We have that $I_0(X') \cap I_1(Y') \subseteq \{X'\}$, and indeed, $\cN^+_q(1,1,0,0) = \binom{1}{0}\binom{1}{1}\binom{2}{0} = 1$.

Assume that the claim in \eqref{clmInter} holds for all $n+t+k < m$. We prove that it is true for $n+t+k = m$. Take sequences $X',Y'$, where $X' \neq Y'$, $X' \in \mathbb{F}_{q}^{n+t-k}, Y' \in \mathbb{F}_q^{n}$, and $d_{e}(X',Y') = t-k+2\ell$, and $n+t+k=m$. Write $X' = aX$ and $Y' = bY$, with $a,b \in \mathbb{F}_q$.

We examine the first symbol. The first case is that $X'$ and $Y'$ start with the same symbol, so that $a = b$. Then, using Lemma~\ref{lem:recsmax}, the induction hypothesis, and the first recursion in Lemma~\ref{lem:recGen}, we write
\begin{align*}
%|I_k(aX) \cap I_t(bY)| &= |I_{k}(X) \cap I_t(Y)| + (q-1) |I_{k-1}(aX) \cap I_{t-1}(aY)| \\
|I_k(aX&) \cap I_t(bY)| \\
&\leq N^+_q(n-1,t,k,\ell) + (q-1)N^+_q(n,t-1,k-1,\ell) \\
&\leq \cN^+_q(n-1,t,k,\ell) + (q-1)\cN^+_q(n,t-1,k-1,\ell) \\
&= \cN^+_q(n,t,k,\ell).
\end{align*}

The other case is $a \neq b$, so that $X'$ and $Y'$ start with different symbols. Now we apply the second result in Lemma~\ref{lem:recsmax}, the induction hypothesis, and the second recursion in Lemma~\ref{lem:recGen}, yielding
\begin{align*}
%|I_k(aX) \cap I_t(bY)| &=  |I_k(X) \cap I_{t-1}(bY)| + |I_{k-1}(aX) \cap I_t(Y)| + (q-2)|I_{k-1}(aX) \cap I_{t-1}(bY)| \\
|I_k&(aX) \cap I_t(bY)|  \\
&\leq N^+_q(n,t-1,k,\ell) + N^+_q(n-1,t,k-1,\ell-1) \\
& \quad + (q-2)N^+_q(n,t-1,k-1,\ell) \\
&\leq \cN^+_q(n,t-1,k,\ell) + \cN^+_q(n-1,t,k-1,\ell-1) \\
& \quad+ (q-2)\cN^+_q(n,t-1,k-1,\ell) \\
&= \cN^+_q(n,t,k,\ell).
\end{align*}

Thus, $N^+_q(n,t,k,\ell) \leq \cN^+_q(n,t,k,\ell)$. Now we show that there exist sequences $X'$,$Y'$ that attain the maximum; that is, we find $X', Y'$ with $|I_k(X') \cap I_t(Y')| = \cN^+_q(n,t,k,\ell)$. This allows us to conclude that $N^+_q(n,t,k,\ell) = \cN^+_q(n,t,k,\ell)$, completing the proof.

We select the sequences \[X'=\underbrace{00\ldots0}_{t-k + n \text{ 0's}} \text{ and } Y' = \underbrace{11\ldots 1}_{\ell \text{ 1's}}\underbrace{00 \ldots 0}_{n-\ell \text{ 0's}}.\] Note of course that we could have selected any sequences with the structure $X'=\underbrace{aa\ldots a  }_{t-k + n \text{ }a\text{'s}}$ and $Y' = \underbrace{bb\ldots b}_{\ell \text{ }b\text{'s}}\underbrace{aa \ldots a}_{n-\ell \text{ }a\text{'s}}$ for $a \neq b$ and $a,b \in \mathbb{F}_q$. 

It is hard to give a direct proof that $|I_k(X') \cap I_t(Y')| = \cN^+_q(n,t,k,\ell)$; instead, we proceed inductively. As we will see, the induction is identical to the previous proof, replacing inequalities with equalities. As before, the induction is on $n+t+k$. The base cases are $n+t+k \in \{1,2\}$. If $n\in\{1,2\}$ and $t=k=\ell=0$, $X'=00$ and $Y'=00$ or $X'=0$ and $Y'=0$. Thus, $|I_0(X') \cap I_0(Y')| = 1 = \cN^+_q(n,0,0,0)$, as desired. If $n=t=1$ and $k=\ell=0$, $X'=00$ and $Y'=0$. Then $|I_0(X') \cap I_1(Y')| = 1 = \cN^+_q(1,1,0,0)$.

Assume that the induction hypothesis holds for $n+t+k < m$; we show it is true for $n+t+k = m$. If $\ell \geq 1$, we apply \eqref{levTrickDiff} to write
\begin{align}
\label{zerosonesexp}
|I_k(X') &\cap I_t(Y')| =  \nonumber \\
&|I_k(X) \cap I_{t-1}(Y')| +|I_{k-1}(X') \cap I_t(Y)|  \nonumber \\ 
&\quad+(q-2)|I_{k-1}(X')\cap I_{t-1}(Y')|,
\end{align} 
where $X=\underbrace{00\ldots0}_{t-k+n-1 \text{ 0's}}$ and $Y = \underbrace{11\ldots 1}_{\ell-1 \text{ 1's}}\underbrace{00 \ldots 0}_{n-\ell \text{ 0's}}$. Note that to produce $X$ from $Y'$ with the fewest operations, we must remove $\ell$ $1$'s and insert $t-k+n-1 - (n-\ell) = \ell + t-k-1$ 0's. Thus, $d_e(X,Y') = t-k+2\ell-1$. A similar calculation gives $d_e(X',Y) = t-k+2\ell-1$. Now we again match arguments: in the first term in \eqref{zerosonesexp}, $Y'$ has length $n$, the insertion arguments are $t-1$ and $k$ and $d_e(X,Y') = t-k+2\ell-1$. Thus, the distance argument satisfies $\ell' = \frac{1}{2}(d_e(X,Y')+k-(t-1)) = \frac{1}{2}(t-k+2\ell-1+k-(t-1))= \ell$. Applying the induction hypothesis, we may thus write $|I_k(X) \cap I_{t-1}(Y')| = \cN^+_q(n,t-1,k,\ell)$. Similarly, for the second term in \eqref{zerosonesexp}, $Y$ has length $n-1$, the insertion arguments are $t$ and $k-1$ while $d_e(X',Y) = t-k+2\ell-1$. The distance argument satisfies $\ell' = \frac{1}{2}(d_e(X',Y)+(k-1)-t) = \frac{1}{2}(t-k+2\ell-1+(k-1)-t) = \ell-1$. Apply the induction hypothesis to write $|I_{k-1}(X') \cap I_t(Y)| = \cN^+_q(n-1,t,k-1,\ell-1)$. For the last term, the arguments are clearly $(n,t-1,k-1,\ell)$. Using the induction hypothesis once more, $|I_{k-1}(X') \cap I_{t-1}(Y')|=\cN^+_q(n,t-1,k-1,\ell)$. Now, using Lemma~\ref{lem:recGen}, we get that
%Since $X$ can be formed from $Y'$ by removing $\ell$ $1$'s and adding $t-k+n-1 - (n-\ell) = \ell + t-k-1$ 0's, $d_e(X,Y') = t-k+2\ell-1$. A similar calculation gives $d_e(X',Y) = t-k+2\ell-1$. Then, using the induction hypothesis and Lemma~\ref{lem:recGen}, we have that 
{\begin{align*}&|I_k(X') \cap I_t(Y')| = \\ 
&\quad \cN^+_q(n,t-1,k,\ell) + \cN^+_q(n-1,t,k-1,\ell-1) \\
&\qquad +(q-2)\cN^+_q(n,t-1,k-1,\ell)  = \cN^+_q(n,t,k,\ell).
\end{align*}}
If $\ell = 0$, $X'$ and $Y'$ both start with $0$ so we apply \eqref{levTrickEqual} to write $|I_k(X') \cap I_t(Y')| = |I_k(X) \cap I_t(Y)| + (q-1)|I_{k-1}(X') \cap I_{t-1}(Y')|.$
In this case too we apply the induction hypothesis and Lemma~\ref{lem:recGen}, giving
\begin{align*}
|I_k&(X') \cap I_t(Y')|   \\
&=\cN^+_q(n-1,t,k,\ell) + (q-1)\cN^+_q(n,t-1,k-1,\ell) \\
&=\cN^+_q(n,t,k,\ell),
\end{align*}
thus completing the proof. To retrieve the formula given by \eqref{maxintersectioneqcase}, take $t=k$ in \eqref{maxintersectionGen}.
\end{IEEEproof}

\subsection{Corollaries}
Specific cases of Theorem~\ref{thm:gen} yield a variety of interesting results. Before we present these results, we require two auxiliary binomial identities. The purpose of these identities is to help simplify the more complex formulas in \eqref{maxintersectionGen} and \eqref{maxintersectioneqcase} for special cases. The proofs are found in the appendix.

\begin{lemma} \label{lem:aux} 
   \begin{enumerate}[label={\arabic*.},wide, labelwidth=!, labelindent=0pt]
       \item For $m\geq 0$, 
		\[\sum_{j=0}^m \binom{2j}{j} \binom{m+j}{2j} (-1)^{m+j} = 1.\] \label{partone}
       \item For $n, m, t, j \geq 0$ and $t+j \geq m$,  
        \[\sum_{i=0}^m \binom{t+j-i}{t+j-m} \binom{n+t}{i} (-1)^{m-i} = \binom{n+m-j-1}{m}. \]
        \label{parttwo}
   \end{enumerate}
\end{lemma}

We are now ready to proceed with our corollaries. We begin by showing that it is possible to recover Levenshtein's formula for $N^+_q(n,t) = \max_{X \neq Y \in \mathbb{F}_q^n} |I_t(X) \cap I_t(Y)|$ by taking $\ell = 1$ in \eqref{maxintersectioneqcase}. In other words, the maximum number of supersequences is attained by taking sequences at the smallest possible ($d_e = 2$) nonzero distance. 
%LEV RESULT L=1--------- --------------------------------------------------------------------------
\begin{cor}[\bf Levenshtein's result for $N^+_q(n,t,\ell=1)$]
\label{cor:levCase}
For $n$ a positive integer and $t$ a non-negative integer,
\begin{align} \label{levCase}
N^+_q(n,t,1) &= N^+_q(n,t) \nonumber \\ 
&= \sum_{i=0}^{t-1} \binom{n+t}{i} (q-1)^i (1-(-1)^{t-i}). 
\end{align}
\end{cor}
\begin{IEEEproof}
{\color{black} Note first that $N^+_q(n,t,1) = N^+_q(n,t)$ follows from Claim~\ref{cl:distance}.} From the first identity in Lemma~\ref{lem:aux}, we have that $\sum_{j=1}^m \binom{2j}{j} \binom{m+j}{2j} (-1)^{m+j} = 1-(-1)^m$. Taking $m=t-i$ yields
\begin{equation} \label{corIdent}
\sum_{j=1}^{t-i} \binom{2j}{j} \binom{t-i+j}{2j} (-1)^{t-i+j} = 1-(-1)^{t-i}.
\end{equation}
If we exchange the sums in $i$ and $j$, we can rewrite the $\ell=1$ case in \eqref{maxintersectioneqcase} as
\begin{align}\label{switchindex}
N^+_q&(n,t,1) = \nonumber \\
&\sum_{i=0}^{t-1} \sum_{j=1}^{t-i} \binom{2j}{j} \binom{t+j-i}{2j} \binom{n+t}{i}(q-1)^i (-1)^{t+j-i}.
\end{align}
Applying \eqref{corIdent}, we have that  
\[N^+_q(n,t,1) = \sum_{i=0}^{t-1} \binom{n+t}{i}(q-1)^i (1-(-1)^{t-i}),\]
as desired.
\end{IEEEproof}

%MINIMUM DISTANCE --------------------------------------------------------------------------
Note that we did not require the distance parameter $\ell$ to be positive in Theorem~\ref{thm:gen}. In fact, $\ell=0$ implies $d_e(X,Y) = 0$, or $X=Y$. In other words, all supersequences of $X$ are ``common'' supersequences (of $X$ and $X$), so we expect $N^+_q(n,t,0)$ to reduce to the formula for the number of supersequences $I_q(n,t)$. Happily, this is the case:
\begin{cor}[\bf $\ell = 0$ case] 
\label{cor:identCase}
For $n$ a positive integer and $t$ a non-negative integer,
\[N^+_q(n,t,0) = I_q(n,t) = \sum_{i=0}^{t} \binom{n+t}{i} (q-1)^i. \]
\end{cor}
\begin{IEEEproof}
We exchange the sums for $i$ and $j$ in \eqref{maxintersectioneqcase} with $\ell=0$. This gives
\begin{align}
\label{switchindex2}
N^+_q&(n,t,0) = \nonumber \\
&\sum_{i=0}^t \sum_{j=0}^{t-i} \binom{2j}{j} \binom{t+j-i}{2j} \binom{n+t}{i}(q-1)^i (-1)^{t+j-i}.
\end{align}
Now set $m=t-i$ in the first part of Lemma~\ref{lem:aux} and apply the result to \eqref{switchindex2}. We get
\[N^+_q(n,t,0) = \sum_{i=0}^t \binom{n+t}{i}(q-1)^i,\]
which is just $I_q(n,t)$.
\end{IEEEproof}

%MAXIMUM DISTANCE -------------------------------------------------------------------------- 
The case $\ell = 0$ represents the minimum distance criterion. We also consider the maximum criterion. Recall that in Theorem~\ref{thm:gen} we required that $k \geq \ell$. What happens if $\ell > k$? In this case, the number of common supersequences is always 0. If a common supersequence $Z$ existed for $X \in \mathbb{F}_q^{n+t-k}$ and $Y \in \mathbb{F}_q^n$ with $d_e(X,Y) = 2\ell + t-k$, we could then produce $Y$ from $X$ with $t+k$ insertions and deletions. However, since $\ell > k$, $t+k < 2\ell + t-k$, a contradiction. This is especially easy to see for equal-length sequences ($t=k$).

The maximum distance with a non-zero number of common supersequences is for $\ell = k$. In that case, the formula in Theorem~\ref{thm:gen} reduces to $\binom{t+k}{k}$. Here, we can even see a direct combinatorial interpretation of the formula. Consider for example $X = 00 \ldots 0$ and $Y = 11 \ldots 1$, where $X$ is made up of $t$ 0's and $Y$ is made up of $k$ 1's. Then, any common supersequence (formed by $t$ insertions into $Y$ and $k$ insertions into $X$) is a sequence with $t$ 0's and $k$ 1's. There are clearly $\binom{t+k}{k}$ such sequences.

%BINARY CASES --------------------------------------------------------------------------
Finally, we introduce a binary simplification of the result in Theorem~\ref{thm:gen}. 
\begin{cor}[\bf Binary Case] 
\label{cor:bincase}
Let $n$ be a positive integer and $t, k, \ell$ be non-negative integers such that $t \geq k \geq \ell$ and $n \geq \ell$. Then, we have the formula
\[N^+_2(n,t,k,\ell) = \sum_{j=\ell}^k \binom{t-k+2j}{j} \binom{n+k-(2j+1)}{k-j}.\]
\end{cor}
\begin{IEEEproof}
We take $m=k-j$ in the second identity of Lemma~\ref{lem:aux}, giving
\[\sum_{i=0}^{k-j} \binom{t+j-i}{t+2j-k} \binom{n+t}{i} (-1)^{k-j-i} = \binom{n+k-2j-1}{k-j}.\]
Now, applying this result, we have 
\begin{align*}
&N^+_2(n,t,k,\ell)   \\
&\enskip= \sum_{j=\ell}^k \sum_{i=0}^{k-j} \binom{t-k+2j}{j}\binom{t+j-i}{t-k+2j} \binom{n+t}{i} (-1)^{k+j-i} \\
&\enskip= \sum_{j=\ell}^k \binom{t-k+2j}{j}  \binom{n+k-2j-1}{k-j}, \end{align*} as desired. 
\end{IEEEproof}

%Next we are ready for a proof of Theorem~\ref{thm:gen}.

\section{Reconstruction in Varshamov-Tenengolts (VT) Codes}
Thus far, we have examined the number of traces required to distinguish between (sufficiently distant) sequences. We answered this question by deriving the $N^+_q(n,t,k,\ell)$ formula. However, this expression represents the worst case. We may wonder whether in some \emph{particular} code the situation is better. That is, we may ask whether the codewords of a code require fewer than $N^+_q(n,t,k,\ell)+1$ traces for reconstruction. 

One major challenge when dealing with such a question is that there are not very many explicit codes correcting a fixed number of insertions and deletions. We will examine the most famous such code, the binary single insertion/deletion-correcting Varshmov-Tenengolts (VT) code \cite{Varshamov, Lev1}: 
\begin{align*}
\cC_{VT}&(n,a) := \\
&\left\{ X=(x_1, \ldots, x_n) \in \mathbb{F}_2^n : \sum_{i=1}^n i \cdot x_i \equiv a \bmod n+1 \right\}.
\end{align*}

We may take any $0 \leq a \leq n$ to form a different code. The VT codes partition the space $\{0,1\}^n$. Since the VT codes correct a single insertion or deletion and have equal-length codewords, for any $E,F \in \cC_{VT}(n,a)$ ($E \neq F$), we have that $d_{e}(E, F) \geq 4$. This represents the case of $\ell=2$ in our notation. Thus we seek to find out whether there exist (unordered) pairs of codewords $\{E,F\}$ with $E,F \in\cC_{VT}(n,a)$ for some particular $a$ such that $|I_t(E)\cap I_t(F)| = N^+_2(n,t,2)$. If such pairs exist for each $a$, we can conclude that the VT codes require the worst-case $N^+_2(n,t,2)+1$ traces for reconstruction. We are motivated to seek codes with similar properties to the VT codes but with fewer traces needed for reconstruction.

In this section, we prove an even stronger result. Not only is there always a pair of codewords achieving the worst case, but there are exponentially many such pairs for each VT code. In particular, we prove that for any $n \geq 7$ and any $a$ with $0 \leq a \leq n$, there exists a set $S_a$ of unordered pairs of elements, where for any element $ \{E,F\} \in S_a$, we have $E, F \in \cC_{VT}(n,a)$, $E \neq F$, $N^{+}_2(n,t,2)=|I_t(E) \cap I_{t}(F)|$ and
\[  |S_a| \geq 2^{n - \lceil \log_2(n) \rceil-3 }. \]
%This implies that for any choice of $a$ there are a large number of pairs of sequences that achieve our upper bound in a VT code of length $n$. 

We first establish some simple claims. We make use of the following notation. Let $V=(v_1, \ldots, v_{n-2}) \in \mathbb{F}_2^{n-2}$ be a sequence of length $n-2$. We write 
\[X(n,p,V):=(v_1,\ldots,v_{p-1},1,1,v_p,\ldots,v_{n-2}) \in \mathbb{F}_2^n,\] 
i.e., $X(n,p,V)$ is a sequence whose components are equal to $V$ in the first $p-1$ positions, followed by $1,1$, then by the remaining $n-p-1$ bits in $V$. Similarly, let \[Z(n,p,V)=(v_1, \ldots, v_{p-1},0,0,v_p, \ldots, v_{n-2}) \in \mathbb{F}_2^n,\] \[Y(n,p,V) = (v_1, \ldots, v_{p-1}, 0, v_{p}, 0, v_{p+1}, \ldots, v_n) \in \mathbb{F}_2^n,\] and  \[W(n,p,V) = (v_1, \ldots, v_{p-1}, 1, v_{p}, 1, v_{p+1}, \ldots, v_n) \in \mathbb{F}_2^n.\] %be the sequence whose components are equal to $V'$ everywhere except in positions $w, w+2$ (clearly $w \leq n-2$).

Before continuing we provide a small example that illustrates the main ideas.

\begin{example} 
We take $V = 100100 \in \mathbb{F}_2^6$. Then, we have that $X(8,4,V) = 10011100$, $Z(8,4,V) = 10000100$. Note that $X(8,4,V)$ and $Z(8,4,V)$ are in the same VT code, $\mathcal{C}_{VT}(8,7)$, since $\sum_{i=1}^8 ix_i = 1+4+5+6  \equiv 7 \bmod 9$ and $\sum_{i=1}^8 iz_i = 1+6 \equiv 7 \bmod 9$. 

\end{example}

Note that we always have that \[d_e(X(n,p,V),Z(n,p,V)) = 4,\] since the two sequences are the same except for the $0,0$ and $1,1$ bits in the middle. Similarly, we have that $d_e(Y(n,p,V),W(n,p,V)) = 4$. Next we show that if we let $p$ be the central position in some $V$, then the resulting sequences $X$ and $Z$ satisfy $|I_t(X)\cap I_t(Z)| = N^+_2(n,t,\ell=2)$ and similarly $|I_t(Y) \cap I_t(W)| = N^+_2(n,t,\ell=2)$. 

\begin{lemma}\label{lem:maxint} For any $n \geq 4$, $t \geq 2$, and $V \in \mathbb{F}_2^{n-2}$,
\begin{align*} N^{+}_2(n,t,2) &= \left| I_t(X(n,\left\lfloor \frac{n}{2}\right\rfloor,V)) \cap I_{t}(Z(n,\left\lfloor \frac{n}{2}\right\rfloor,V)) \right|  \\
&=\left| I_t(Y(n,\left\lfloor \frac{n}{2} \right\rfloor,V)) \cap I_{t}(W(n,\left\lfloor \frac{n}{2}\right\rfloor,V)) \right|. 
\end{align*}
\end{lemma}
\begin{IEEEproof} We show that $N^{+}_2(n,t,2) = | I_t(X(n, \lfloor \frac{n}{2} \rfloor,V)) \cap I_{t}(Z(n,\lfloor \frac{n}{2}\rfloor,V)) |$. Just as in our earlier proofs, we use induction; this time on $n+t$. Since $n\geq 4$ and $t\geq 2$, we have $n+t=6$ as our base case. It can be verified exhaustively that for any $V \in \mathbb{F}_2^2$, $N^{+}_2(4,2,2)=| I_2(X(4,2,V)) \cap I_{2}(Z(4,2,V)) | = \sum_{j=\ell}^t \binom{2j}{j} \binom{n+t-(2j+1)}{t-j} = \binom{4}{2}\binom{10-5}{2-2} = 6$, as desired.

Suppose that $N^{+}_2(n,t,2) = | I_t(X(n, \lfloor \frac{n}{2} \rfloor,V)) \cap I_{t}(Z(n,\lfloor \frac{n}{2} \rfloor,V)) |$ for all $n+t < m$ and consider the case where $n+t=m$. Suppose that $n$ is even. The case where $n$ is odd can be proven using similar arguments. Let $V' \in \mathbb{F}_2^{n-3}$ be the sequence obtained by deleting the first bit from $V$. Since $X(n, \frac{n}{2},V)$ and $Z(n,\frac{n}{2},V)$ start with the same bit, we can use \eqref{levTrickEqual} in Claim~\ref{levTrick} to write
\begin{align*}
&\left|I_t(X(n, \frac{n}{2},V)) \cap I_{t}(Z(n,\frac{n}{2},V)) \right| = \\
&\enskip \left| I_t(X(n-1, \left\lfloor \frac{n-1}{2} \right\rfloor,V' )) \cap I_{t}(Z(n-1,\left\lfloor \frac{n-1}{2} \right\rfloor,V')) \right| \\
&\quad +  \left| I_{t-1}(X(n, \frac{n}{2},V )) \cap I_{t-1}(Z(n,\frac{n}{2},V))  \right|.
\end{align*}
Applying the inductive hypothesis, $|  I_t(X(n-1, \lfloor \frac{n-1}{2} \rfloor,V')) \cap I_{t}(Z(n-1, \lfloor \frac{n-1}{2} \rfloor,V')) | = N^{+}_2(n-1,t,2)$ and $|  I_{t-1}(X(n, \frac{n}{2},V )) \cap I_{t-1}(Z(n, \frac{n}{2}, V)) | = N^{+}_2(n,t-1,2)$. From Lemma~\ref{lem:recGen} {\color{black} and the fact that $N^+_q(n,t,\ell) = \cN^+_q(n,t,\ell)$}, $N^{+}_2(n-1, t, 2) + N^{+}_2(n,t-1,2) = N^{+}_2(n,t,2)$ so that $ N^{+}_2(n,t,2) = | I_t(X(n,\lfloor \frac{n}{2}\rfloor,V)) \cap I_{t}(Z(n, \lfloor \frac{n}{2} \rfloor, V)) |$ as desired. The expression $N^{+}_2(n,t,2)= | I_t(Y(n,\lfloor \frac{n}{2} \rfloor,V)) \cap I_{t}(W(n, \lfloor \frac{n}{2} \rfloor, V)) |$ can be proven using nearly identical logic.
\end{IEEEproof}

Our eventual goal is to find codeword pairs that are in a particular VT code and achieve the worst-case number of common supersequences. These pairs must satisfy the checksum constraint that defines the VT code. To ensure this, we need to control certain positions in these codewords. We use a function $FP$ that takes as an argument an integer $n$ and returns a subset of integers (related to the positions that we control) of size at most $\lceil \log (n) \rceil + 1$. The set returned by $FP$ is defined iteratively:
\begin{enumerate}
\item Initialize $T = \{1\}$.
\item Let $t = \sum_{i \in T} i$.
\begin{enumerate}
\item If $t \geq n$, then define $FP(n) = T$ and stop.
\item If $n$ is even and $t+1 \in \{ \frac{n}{2}, \frac{n}{2} + 1\}$, then let $T = T \cup \{\frac{n}{2} -1\}$ and go back to step 2).
\item If $n$ is odd and $t+1 = \lfloor \frac{n}{2} \rfloor$, then let $T = T \cup  \{\lfloor \frac{n}{2} \rfloor-1\}$ and go back to step 2). If $n$ is odd and $t+1= \lfloor \frac{n}{2} \rfloor + 2$, then let $T = T \cup \{ \lfloor \frac{n}{2} \rfloor + 1\}$ and go back to step 2).
\item Set $T = T \cup \{t+1\}$, and go back to step 2).
\end{enumerate}
\end{enumerate}

We illustrate the previous procedure via an example.

\begin{example} Suppose $n=16$. Then, after step 1) of the procedure to compute $FP(16)$, we have $T=\{1\}$. Next, $t=1$ and we go to step 2-d) and get $T=\{1,2\}$. Afterwards, we again go to step 2-d) and have $T=\{1,2,4\}$. At this point $t=7$ so that $t+1 = \frac{n}{2}$. Then from step 2-b), $T=\{1,2,4,7\}$. Next, $t=14$ and we go to step 2-d) again so that $T = \{1,2,4,7,15\}$. Finally, we reach step 2-a), and the procedure stops so that $FP(16) = T = \{1,2,4,7,15\}$. Notice that $|FP(16)| = \log_2(16) + 1=5$.  \end{example}

The idea of the algorithm producing the output of $FP(n)$ is to include positions that are roughly powers of 2 while avoiding certain central positions based on the parity of $n$. The reason for the avoidance is that sequences such as $X(n, \lfloor \frac{n}{2} \rfloor,V')$ have these positions already fixed and we therefore cannot control them in order to ensure that the output sequence is in a particular VT code. The remaining positions, however, form a basis, (that is, a linear combination of them produces any $a$ with $0 \leq a \leq n$ modulo $n+1$) so that we can control their output in the checksum to fix the $a$ VT code parameter. 

{\color{black} The algorithm enables us to find the following result. We defer the proof to the appendix.}

%\begin{lemma}\label{lem:Fprop} For any $n \geq 4$, and integer $m \leq n$, there exists a subset $T' \subseteq FP(n)$ where $\sum_{i \in T'} i = m $ and $\lceil \log_2(n) \rceil \leq |FP(n)| \leq \lceil \log_2(n) \rceil + 1$. Furthermore, if $n$ is even, we have $\{ \frac{n}{2}, \frac{n}{2} + 1\} \not \in FP(n)$ and if $n$ is odd, then $\{ \lfloor \frac{n}{2} \rfloor,\lfloor \frac{n}{2} \rfloor+2 \} \not \in FP(n)$.
\begin{lemma}\label{lem:Fprop} For any integer $n \geq 7$, and integer $0 \leq m \leq n$, there exists a subset $T' \subseteq FP(n)$ where $\sum_{i \in T'} i = m $. In addition, $\lceil \log_2(n) \rceil \leq |FP(n)| \leq \lceil \log_2(n) \rceil + 1$. Furthermore, if $n$ is even, we have $\{ \frac{n}{2}, \frac{n}{2} + 1\} \not \in FP(n)$ and if $n$ is odd, then $\{ \lfloor \frac{n}{2} \rfloor,\lfloor \frac{n}{2} \rfloor+2 \} \not \in FP(n)$.
\end{lemma}
%\begin{IEEEproof} We should probably prove this. Should just be an inductive proof but nonetheless a pain.
%\end{IEEEproof}

The idea in Lemma~\ref{lem:Fprop} is that a linear combination of at most $\lceil \log_2(n) \rceil$ numbers in $[n]$ can generate any $m$ for $0 \leq m \leq n$. These numbers are precisely those returned by the function $FP(n)$. Therefore, we can force a sequence of length $n$ to be in any particular VT code by controlling the sequence components at these positions. The subset $T'$ represents those positions where we will place a 1 in the codeword, while the positions given by $FP(n) \setminus T'$ will be set to 0. The idea is explained in further detail in the following proof of the main result of this section.

%We are now ready to proceed with the main result of this section.

\begin{theorem} For any $n \geq 7$ and $a \in \mathbb{Z}_{n+1}$, there exists a set of unordered pairs $S_{a}$ where for any pair $\{ E, F \} \in S_{a}$, we have $E, F \in \cC_{VT}(n,a)$, $E \neq F$, $N^{+}_2(n,t,2)=|I_t(E) \cap I_{t}(F)|$, and  
$$  |S_{a}| \geq 2^{n - \lceil \log_2(n) \rceil-3 }. $$
 \end{theorem}
\begin{IEEEproof} 
We start by counting the number of different ways to form a pair $\{E, F\} \in S_{a}$. First, consider the case where $n$ is even. Let $U = FP(n) \cup \frac{n}{2} \cup (\frac{n}{2}+1)$ be a set of positions in our codewords. We can select the remaining positions, that is, $[n]\setminus U$, freely and still form a codeword. Therefore, we have $2^{n-|FP(n)|-2}$ choices for these positions. Next, let us say that $\sum_{k \in [n] \setminus U} k e_k \equiv c \bmod n+1$. Then, using Lemma~\ref{lem:Fprop}, we fix the components of $E$ at positions in the set $FP(n)$ so that $\sum_{k \in FP(n)} k e_k \equiv a - c \bmod n+1$. % From Lemma~\ref{lem:Fprop}, it is possible to assign values to $E$ so that $\sum_{k \in FP(n)} e_k \equiv a - c \bmod n+1$. 
Finally let $(e_{\frac{n}{2}}, e_{\frac{n}{2} + 1})=(0,0)$. Thus, we may write $E = Z(n,\frac{n}{2},E')$ for some vector $E' \in \mathbb{F}_2^{n-2}$. Notice that $E \in \cC_{VT}(n,a)$ since $\sum_{k \in [n]} k e_k = \sum_{k \in [n] \setminus U} k e_k + \sum_{k \in FP(n)} k e_k + \frac{n}{2} e_{\frac{n}{2}} + (\frac{n}{2}+1) e_{\frac{n}{2}+1} \equiv a \bmod n+1$ as desired.

%We begin by fixing the value of $X'=(x_1, \ldots, x_n)$ except in positions from the set $U = FP(n) \cup \frac{n}{2} \cup (\frac{n}{2}+1)$. Recall that, as a consequence of Lemma~\ref{lem:Fprop}, we have $\{ \frac{n}{2}, \frac{n}{2} + 1\} \not \in F(n)$. Thus, there are $2^{n-|FP(n)|-2}$ ways to choose the element $X'$. 

%Suppose $\sum_{k \in [n] \setminus U} x_k \equiv c \bmod (n+1)$. Then, we fix the components of $X'$ at positions in the set $FP(n)$ so that $\sum_{k \in FP(n)} x_k \equiv a - c \bmod (n+1)$. From Lemma~\ref{lem:Fprop}, it is possible to assign values to $X'$ so that $\sum_{k \in F(n)} x_k \equiv a - c \bmod (n+1)$. Finally let $(x_{\frac{n}{2}}, x_{\frac{n}{2} + 1})=(0,0)$. Notice that $X' \in \cC_{VT}(a)$ since $\sum_{k \in [n]} x_k = \sum_{k \in [n] \setminus U} x_k + \sum_{k \in F(n)} x_k + x_{\frac{n}{2}} + x_{\frac{n}{2}+1} \equiv a \bmod (n+1)$ as desired. 

Let $F = (e_1, \ldots, e_{\frac{n}{2}-1}, 1,1, e_{\frac{n}{2}+2}, \ldots, e_n)=(f_1, \ldots, f_n)$, so that $F = X(n,\frac{n}{2},E')$. Then, $F \in \cC_{VT}(n,a)$, since $\sum_{k \in [n]} k f_k = \sum_{k \in [n]} k e_k + \frac{n}{2} + (\frac{n}{2} + 1) \equiv a \bmod (n+1).$ Then, from Lemma~\ref{lem:maxint}, $|I_t(E) \cap I_{t}(F)| = N^{+}_2(n,t,2)$. Thus, $|S_a| \geq 2^{n-|FP(n)|-2} \geq 2^{n - \lceil \log_2(n) \rceil - 3}$.

Next we examine the case of $n$ odd. We proceed in the same manner as before except that we first select the values of $E$ except in positions from the set $U= FP(n) \cup \lfloor \frac{n}{2} \rfloor \cup ( \lfloor \frac{n}{2} \rfloor +2)$. Afterwards we assign values to components in $E$ whose indices belong to set $FP(n)$ in such a way that $E \in \cC_{VT}(n,a)$. Finally $e_{ \lfloor \frac{n}{2} \rfloor}, e_{\lfloor \frac{n}{2} \rfloor + 2}$ are both set to zero so that $E = Y(n,\lfloor \frac{n}{2} \rfloor,E')$. Next, $F=(f_1, \ldots, f_n)$ is set to be equal to $E$ except that $f_{\lfloor \frac{n}{2} \rfloor}=f_{\lfloor \frac{n}{2} \rfloor +2} = 1$. Thus $F = W(n,\lfloor \frac{n}{2} \rfloor, E')$. Using the same arguments as in the previous paragraph, it can be shown that $E,F \in \cC_{VT}(n,a)$. Furthermore, from Lemma~\ref{lem:maxint}, $|I_t(E) \cap I_{t}(F)| = N^{+}_2(n,t,2)$ and thus $|S_a| \geq  2^{n - \lceil \log_2(n) \rceil - 3}$.
\end{IEEEproof}

Next, we consider other channels and applications of our main results.

\section{Other Channels and Applications}
Thus far we have only been concerned with insertion channels. It is reasonable to ask what occurs in the cases of deletion or mixed insertion/deletion channels. It is not too surprising that finding expressions for similar problems for these channels is much harder: the deletion channel is much less symmetric compared to the insertion channel, and the insertion/deletion channel deals with the challenges of both. {\color{black} We also apply the results (derived for an adversarial channel) to a probabilistic insertion channel.} %We provide a few specific results while leaving the general questions open for further study. 

\subsection{Deletion Channel}
Levenshtein examined exact reconstruction for deletion channels in \cite{Lev4}. He defined $N^-_q(n,t) := \max_{X,Z \in \mathbb{F}_q^n, X \neq Z} |D_t(X) \cap D_t(Z)|$ and showed that
\[N^-_q(n,t) = {\color{black} \left( \sum_{i=1}^{q-1} D_q(n-i-1,t-i) \right)} + D_q(n-2,t-1).\]
Here, $D_q(n,t)$ is the maximum size of the deletion ball $D_t(X)$ for some $X \in \mathbb{F}_q^n$. It is known that $D_q(n,t)$ satisfies the recursion $D_q(n,t) = \sum_{i=0}^t \binom{n-t}{i} D_{q-1}(t,t-i),$ where $D_1(n,t) = 1$ if $n \geq t \geq 0$ and $D_q(n,t)=0$ otherwise \cite{Calabi}. It is not hard to see that $D_2(n,t) = \sum_{i=0}^t \binom{n-t}{i}$. This enables us to write that the maximum number of common subsequences in the binary case is given by
\[ N^-_2(n,t) = 2\sum_{i=0}^{t-1}\binom{n-t-1}{i} .\]
Just as before, we may ask what happens to the number of sequences required for reconstruction if we select the original sequences from an insertion/deletion-correcting code. We can analogously define
\[N_q^-(n,t,\ell) = \max_{\substack{X,Z \in \mathbb{F}_q^n\\ d_e(X,Z) \geq 2\ell}} |D_t(X) \cap D_t(Z)|.\]
Few results are known regarding $N_q^-(n,t,\ell)$. The work \cite{RyanDel} is dedicated to the $\ell=2, q=2$ case (corresponding to VT codes). The authors showed that for $t \leq n/2$,
\begin{align*}
N^-_2(n,t,2) =& 2D_2(n-4,t-2)+2D_2(n-5,t-2)\\
&+2D_2(n-7,t-2) + D_2(n-6,t-3)\\
&+D_2(n-7,t-3).
\end{align*}
Our contribution consists of removing the reliance on recursions from this formula, yielding the exact expression
\begin{align*}
N^-_2(n,t,2) &= 2D_2(n-2,t-1) - 2\binom{n-t-3}{t-1}\\
&\qquad \qquad-\binom{n-t-4}{t-3}-\binom{n-t-5}{t-3} \\
&= 2\sum_{i=0}^{t-1}\binom{n-t-1}{i} -2\binom{n-t-3}{t-1} \\
&\qquad\qquad -\binom{n-t-4}{t-3}-\binom{n-t-5}{t-3}.
\end{align*}
The proof is an easy induction.

\subsection{Insertion/Deletion Channel}
What about the case of insertion/deletion channels? In general, this problem is quite hard, since even the sizes of $t$-insertion/$t$-deletion balls are not known beyond trivial cases. Let us slightly abuse notation as follows: given a set $S \subseteq \mathbb{F}_q^n$, we write $I_t(S)$ and $D_t(S)$ for $\cup_{X \in S} I_t(X)$ and $\cup_{X \in S} D_t(X)$, respectively. Then, the $t$-insertion/$t$-deletion ball centered $X$ may be written $B_t(X) := I_t(D_t(X))$. 

Since the general version of the problem seems intractable, in this subsection we focus on providing a lower bound on the number of distinct distorted sequences (resulting from an insertion/deletion channel) required to reconstruct a binary sequence $X$. Specifically, we are interested in a lower bound on $N_H(\mathbb{F}_2^n,2t)$, {\color{black} defined as in \eqref{definter},} where $H$ is the set of single symbol insertions and deletions. Note that here, we specifically require the $2t$ argument to imply $t$ insertions and $t$ deletions. We can write, in general, that

\[N_H(\mathbb{F}_2^n,2t) = \max_{\substack{X,Z \in \mathbb{F}_2^n \\ X \neq Z}} |I_t(D_t(X)) \cap I_t(D_t(Z))|.\]

We provide a lower bound on $N_H(\mathbb{F}_2^n,2t)$ by computing the number of common distorted sequences in one particular (and non-trivial) case. This is the case of the so-called binary circular string $C_n = \underbrace{0101\ldots}_{n \text{ bits }}$ (or $\underbrace{10101\ldots}_{n \text{ bits }}$). This string is particularly interesting; in \cite{Calabi} it is shown that\footnote{There are no expressions for $|D_t(X)|$ for general $t$; however, the minimal, maximal, and average values are known. The tightest known bounds on $|D_t(X)|$ are found in \cite{Liron}.} \[C_n = \arg \max_{X \in \mathbb{F}_2^n} |D_t(X)|.\]

We begin by evaluating the size of the {\color{black}ball} centered at $C_n$, $B_t(C_n) = I_t(D_t(C_n))$. 

\begin{theorem} \label{altStringDegree}
The size of the {\color{black} the ball around} the binary circular string $C_n \in \mathbb{F}_2^n$ is given by
\begin{equation}\label{degAn} 
|B_t(C_n)| = |I_t(D_t(C_n))| = \sum_{i=0}^{2t} \binom{n}{i} .\end{equation}
\end{theorem}

Before we proceed with the proof of Theorem~\ref{altStringDegree}, we comment on this result. Since $C_n$ is known to maximize the deletion ball size $|D_t(X)|$, we may ask whether the string $C_n$ also maximizes the insertion/deletion {\color{black} ball} size. Surprisingly, this is not the case. Although  $|B_t(C_n)|$ is quite large and in certain cases is in fact maximal, the string $X = 00110011\ldots$ generally yields a larger ball size. More details on which strings maximize {\color{black}such ball sizes} can be found in \cite{Cullina}.

We use the following lemma in the proof of Theorem~\ref{altStringDegree}:
\begin{lemma} \label{lcirc}
Let $n,t$ be positive integers with $n \geq 2t$. Let $C_{n-2t}$ be the substring formed by the first $n-2t$ bits of the circular string $C_n$. Then, the $t$-deletion ball centered at $C_{n}$ is exactly the $t$-insertion ball centered at $C_{n-2t}$. That is,
\begin{equation}  \label{alterneq}
D_t(C_{n}) = I_t(C_{n-2t}).
\end{equation}
\end{lemma}
\begin{IEEEproof}
First, observe that $C_{n-2t}$ begins and ends with the same bit as $C_{n}$. We will show the result by induction on $n$.

The base case is $n=2t$. Here, $C_{n-2t}$ is the empty string, and the right hand side in \eqref{alterneq} is just $\mathbb{F}_{2}^t$, the set of all binary sequences of length $t$. It is easy to see that this set is equal to $D_t(C_{n}) = D_t(C_{2t}) = D_t(01 \ldots 01)$ (or $D_t(10\ldots 10)$). We may delete either the 0 or the 1 in all of the $t$ consecutive $01$ (or $10$) pairs in order to produce any sequence of length $t$. This establishes the base case.

Now, we assume that $D_{t'}(C_{m}) = I_{t'}(C_{m-2t})$ for all $t' \leq t$ and $m$ satisfying $2t \leq m\leq n$. The cases of $t' < t$ follow from the $t$ case by deleting and inserting identical elements. We examine $D_t(C_{n+1})$ with the goal of showing that it is identical to $I_{t}(C_{n+1-2t})$. We take the last bit of $C_{n+1}$ (and thus, of $C_{n+1-2t})$ to be $1$, without loss of generality. Consider some $X \in D_t(C_{n+1})$ so that $X$ ends in exactly $k$ consecutive 0s, with $0 \leq k \leq t$. We show that $X \in I_t(C_{n+1-2t})$.

If $k=0$, $X$ ends in 1, like $C_{n+1}$. In this case, $X = Y1$ where $Y$ has length $n-t$. Then, $Y$ may be produced by $t$ deletions in the string $C_n$, which is itself the first $n$ bits of $C_{n+1}$. Thus, $Y \in D_t(C_n)$. By the induction hypothesis, $D_t(C_n) = I_t(C_{n-2t})$, so $Y \in I_t(C_{n-2t})$. Then, $Y$ can be produced by $t$ insertions to $C_{n-2t}$, so, since $C_{n-2t+1}$ ends in 1, indeed $X \in I_{t}(C_{n-2t+1})$.

 If $0< k \leq t$, $X$ ends with the substring $1\underbrace{00\ldots 0}_{k\text{ 0's}}$. In fact, we may write $X = Y \underbrace{00\ldots 0}_{k\text{ 0's}}$ for some string $Y$ of length $(n+1-t-k)$. $X$ results from the deletion of the last $k$ 1's from $C_{n+1}$, and the deletion of an additional $t-k$ elements from the first $n+1-2k$ bits of $C_{n+1}$, which themselves form $C_{n+1-2k}$. That is, $Y \in D_{t-k}(C_{n+1-2k})$. Applying the induction hypothesis, $Y$ is in the set $I_{t-k}(C_{n+1-2k-2(t-k)}) = I_{t-k}(C_{n+1-2t})$. Then, clearly $X \in I_t(C_{n+1-2t})$, as we may use the remaining $k$ insertions to add $k$ 0s to the end of $Y$ to produce $X$. We conclude that $D_t(C_{n+1}) \subseteq I_t(C_{n+1-2t}).$
 
 The other direction is essentially identical. Take $Z \in I_t(C_{n+1-2t})$. If $Z$ ends in $1$, then $Z = Y1$ where $Y$ may be formed by $t$ insertions into $C_{n-2t}$. By the induction hypothesis, $Y \in D_{t}(C_{n})$, and since $C_{n+1}$ ends in 1, we have that $Z \in D_t (C_{n+1}).$ If $Z$ ends in exactly $k$ 0s, ($1 \leq k \leq t$) then $Z =Y \underbrace{00\ldots 0}_{k\text{ 0's}}$ for some $Y$ of length $n+1-t-k$. Then, $Y$ can be formed by $t-k$ insertions into $C_{n+1-2t}$. By the induction hypothesis, $Y \in D_{t-k}(C_{n+1-2t+2(t-k)}) = D_{t-k}(C_{n+1-2k})$. Then, $Z \in D_t(C_{n+1})$, since we may use the remaining $k$ deletions to delete the last $k$ 1s in $C_{n+1}$. With this, $I_t(C_{n-2t+1}) \subseteq D_t(C_{n+1})$.

Thus, $D_t(C_{n+1}) = I_t(C_{n-2t+1})$, and we are done. 
\end{IEEEproof}

Theorem~\ref{altStringDegree} follows almost immediately from Lemma~\ref{lcirc}:
\begin{IEEEproof}
Let $n \geq 2t$. According to Lemma~\ref{lcirc}, $D_t(C_{n}) = I_t(C_{n-2t})$. Then, we have that 
\begin{align*}
|B_t(C_n)| &= |\cup_{X \in D_t(C_{n})} I_t(X)| \\
&= |\cup_{X \in I_t(C_{n-2t})} I_t(X)| \\
&= |I_{2t} (C_{n-2t})|  \\
&=  \sum_{i=0}^{2t} \binom{n}{i},
\end{align*} where in the last step, we used the formula {\color{black} \eqref{eq:insform} for the number of supersequences formed by $2t$ insertions}.
The remaining cases for $n <2t$ are identical to the base case $n=2t$ in the proof of Lemma~\ref{lcirc}. Here too, $D_t(C_{n}) = F^{n-t}_2$,  so that $\cup_{X \in D_t(C_{n})} I_t(X) = \cup_{X \in F^{n-t}_2} I_t(X)$, implying that $|B_t(C_n)| = 2^n$.
\end{IEEEproof}

Theorem~\ref{altStringDegree} is interesting, as in general it is very difficult to compute the exact ball size $|B_t(X)|$ for any non-trivial $X$ (such as any sequence that is not made up of all 0's or all 1's) or $t>1$. The underlying symmetries for the circular string enable us to give this exact expression. We remark that Lemma~\ref{lcirc} also yields an alternative way to compute the size of $D_t(C_n)$ \cite{Calabi}.

Now we return to the problem of common distorted sequences. Recall that we are interested in computing $|I_t(D_t(X)) \cap I_t(D_t(Z))|$ for at least some non-trivial $X,Z \in \mathbb{F}_2^n$. Let us take $X = C_n = 10101\ldots$ and $Z = C_n' = 010101 \ldots$. Note that $d_e(C_n, C_n') = 2$, since we need only take the leading 1 in $C_n$ and move it to the end to reproduce $Z$. Now, we have that
\begin{align*}
|I_t(D_t(C_n)) &\cap I_t(D_t(C_n'))| \\
&= |I_t(I_t(C_{n-2t})) \cap I_t(I_t(C_{n-2t}'))| \\
&= |I_{2t}(C_{n-2t}) \cap I_{2t}(C_{n-2t}')| \\
&= N^+_2(n-2t,2t,1) \\
&= 2\sum_{i=0}^{2t-1} \binom{n}{i} .
\end{align*}

The equality (rather than inequality) in the third step is easy to check. Therefore, we have our desired bound on $N_H(\mathbb{F}_2^n,2t)$:
\begin{equation*}N_H(\mathbb{F}_2^n,2t) \geq 2\sum_{i=0}^{2t-1} \binom{n}{i} .\end{equation*}

The important idea here is to replace deletions in our insertion/deletion channel with insertions. This idea is often useful when computing sizes of insertion/deletion balls, since deletions are much more difficult to deal with. Note that we can use a similar idea to compute the number of common distorted sequences for some other cases. For example, if we let $Z = C_n$, but take $X = 00\ldots 0$, we have that $I_t(D_t(0 \ldots 0)) = I_t(0\ldots 0) = I_2(n-t,t)$, which yields $|  I_t(D_t(0\ldots 0)) \cap I_t(D_t(C_n)) | = |  I_t(0 \ldots 0) \cap I_{2t}(C_{n-2t})| \leq N^+_2(n-2t,2t,t, \frac{1}{2}(\lfloor \frac{n-2t}{2} \rfloor-t))$. A number of other similar expressions can be computed.%= \sum_{j=\lfloor \frac{t}{2} \rfloor}^t \binom{2j}{j}\binom{n - (2j+1)}{t-j}$. A number of other similar expressions can be computed.

Having examined the deletion and insertion/deletion channels, we also consider applying our results to non-adversarial channels.

{\color{black}
\subsection{Application to Probabilistic Channel}
Thus far, we have entirely focused on adversarial channels. However, there are many practical communication scenarios that can be modeled with probabilistic synchronization channels (e.g., underwater communication \cite{underwater} and file synchronization \cite{synch}, \cite{FredSynch}). Our results can be applied to such channels as well. We focus on the standard probabilistic insertion channel shown in Fig.~\ref{markov_insert}. Let $\bf x$ be a sequence in $\{0,1,\ldots, q-1\}^n$ being transmitted through the channel. The insertion process is modeled as a Markov chain. For each symbol $x_j$ for $1 \leq j \leq n$, there is an associated state which we call $s_j$. The initial state is $s_1$. At state $s_j$, $1 \leq j \leq n$, there is an option to insert a symbol in $\{0,1,\ldots, q-1\}$ and return to state $s_j$, with some probability $p$ ($0 \leq p < 1$), or to transmit symbol $x_j$, with probability $1-p$. After transmitting $x_j$ (from the original sequence ${\bf x}$), the process continues to state $s_{j+1}$ if $j<n$ or terminates if $j=n$.

\begin{figure}
\centering
\includegraphics[width=\linewidth]{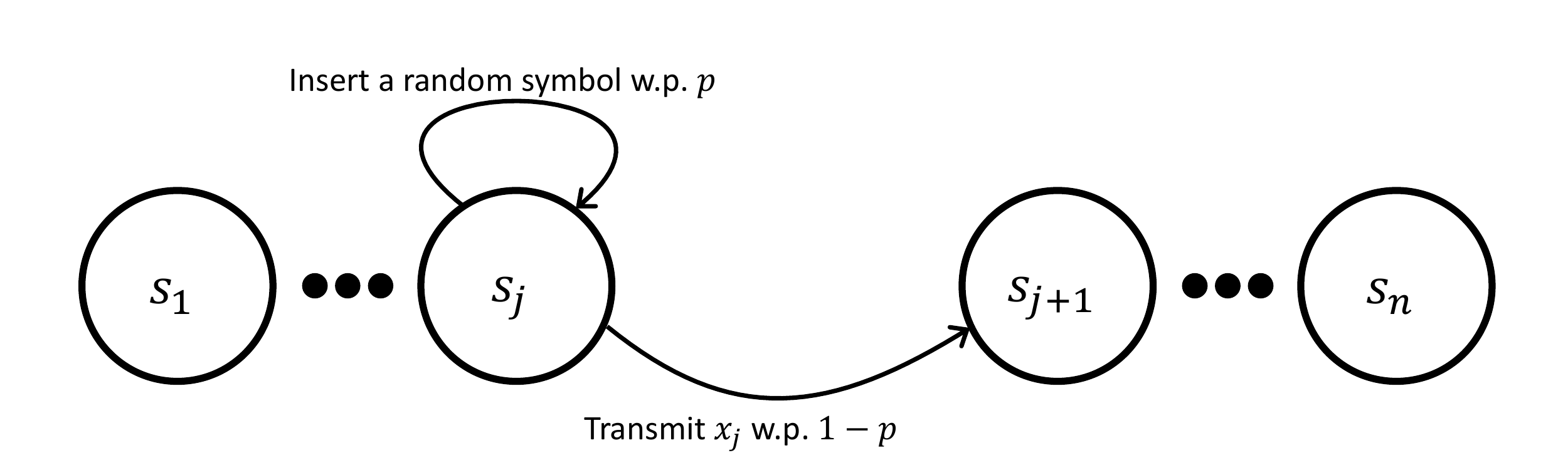}
\caption{Probabilistic insertion channel.}
\label{markov_insert}
\end{figure}

Let $\mathcal{C} \subseteq \{0,1,\ldots,q-1\}^n$ be a code with minimum edit distance $2\ell$ and length $n$ and suppose ${\bf x} \in \mathcal{C}$. In the probabilistic scenario, we repeatedly transmit ${\bf x}$ through the channel and accumulate outputs until we can decode/reconstruct $\bf x$. The outputs from repeated transmissions of $\bf x$ will be of varying lengths. Any channel output of length in $\{n,n+1, \ldots, n+\ell-1\}$ is decodable (with code $\mathcal{C}$'s decoder). If $N_q(n,v,\ell)+1$ outputs of length $(n+v)$ (where $v\geq \ell$) are received by repeated transmissions before an output with length in $\{n,n+1, \ldots, n+\ell-1\}$ is received, we will uniquely reconstruct (rather than decode) $\bf x$. This scenario will occur if the probability of insertion $p$ is sufficiently high. The forthcoming analysis will show such cases where reconstruction occurs before decoding. This idea shows the applicability of our result to probabilistic channels.

%This scenario will occur if the probability of insertion $p$ is sufficiently high, the probability of receiving an output of length in $\{n,n+1, \ldots, n+\ell-1\}$ is low compared to that of receiving an output of length $n+v$ (where $v\geq \ell$).  The forthcoming analysis will show such cases where reconstruction occurs before decoding. This idea shows the applicability of our result to probabilistic channels.

We let $p_i$ be the probability that a particular channel output has been affected by exactly $i$ insertions for $i \geq 0$. Let $T_i$ be a threshold that counts the required number of length $(n+i)$ outputs for decoding/reconstruction of $\bf x$. We seek to answer the following question: what is the average required number of transmissions to collect $T_i$ channel outputs of length $(n+i)$? In particular, for what value of $i$ is this quantity minimized?

%on average, how many total transmissions are needed for some sequences of a particular length to yield enough channel outputs to correctly decode, either by the code's decoder (for short lengths) or by reconstruction (for longer lengths)? . We now state the question as follows: what is the average required number of transmissions to collect $T_i$ channel outputs of length $n+i$? In particular, for what value of $i$ is this quantity minimized?

First we compute a formula for the $T_i$'s. The code $\mathcal{C}$ has minimum distance $2\ell$, so that it can correct up to $\ell-1$ insertions. As a result, a single output of length $n, n+1, \ldots, n+\ell-1$ is sufficient to decode. If the outputs have length $n+v$ for $v \geq \ell$, we need $N^+_q(n,t,\ell)+1$ outputs for reconstruction. Thus,
  \[
T_i = 
  \begin{cases}
    1, & \text{for } 0 \leq i \leq \ell-1, \\
    N^+_q(n,t,\ell)+1, & \text{for } \ell \leq i. \\
  \end{cases}
  \]

Next we can compute $p_i$ as follows. There are $i$ insertions. Since the channel allows for multiple insertions to take place at each state, we select $r$ ($1 \leq r \leq n$) states for the $i$ insertions. The total number of ways to distribute $i$ objects into $r$ buckets with at least one object per bucket is $\binom{i-1}{r-1}$. There are $\binom{n}{r}$ symbols selected for insertion. Therefore, there are $\binom{n}{r}\binom{i-1}{r-1}$ ways to perform the insertions. Since there are $i$ insertions, we have a factor of $p^i$. From the channel procedure, to move on from each state, we need a factor of $(1-p)$, for a total of $(1-p)^n$ for $n$ states. Putting it all together, $p_0 = (1-p)^n$, while for $i \geq 1$, 
\[p_i = \sum_{r=1}^{\min(n,i)} \binom{n}{r} \binom{i-1}{r-1} p^i (1-p)^n.\]

The number of transmissions required to collect $T_i$ channel outputs of length $(n+i)$ follows a negative binomial distribution. For simplicity, we ignore the possibility of identical channel outputs of the same length. The mean number of trials to generate $f$ successes is given by $\frac{f}{p}$. Thus, we have that the expected number of transmissions is 
\[E_i = \frac{T_i}{p_i},\]
or
\[E_i = 
  \begin{cases}
%   1/\left(\sum_{r=1}^{\min(n,i)} \binom{n}{r} \binom{i-1}{r-1} p^i (1-p)^n \right) , \\
%    \pushright{ \text{for } 0 \leq i \leq \ell-1, \qquad} \\
%   (N^+_q(n,t,\ell)+1) /\left(\sum_{r=1}^{\min(n,i)} \binom{n}{r} \binom{i-1}{r-1} p^i (1-p)^n \right), \\
%   \pushright{\text{for } \ell \leq i.\qquad} \\
    \cfrac{1}{\sum_{r=1}^{\min(n,i)} \binom{n}{r} \binom{i-1}{r-1} p^i (1-p)^n } , & \text{for } 0 \leq i \leq \ell-1, \\
    \cfrac{N^+_q(n,t,\ell)+1}{\sum_{r=1}^{\min(n,i)} \binom{n}{r} \binom{i-1}{r-1} p^i (1-p)^n }, & \text{for } \ell \leq i. \\
  \end{cases}
\]

It is possible to numerically compute $n+\arg \min_i E_i$, the expected length of the output sequence that will first allow us to decode/reconstruct. Indeed, it can be the case that reconstruction will be performed sooner than decoding. Consider, for example, $\mathcal{C}$ a VT code of length $n=500$ and distance $2\ell=4$. Set $p=0.3$. Then, $E_2$ transmissions are required on average for the $\binom{2\times2}{2}+1=7$ outputs of length $n+2$ needed for reconstruction. We need $E_1$ or $E_0$ expected transmissions to decode. In this case, $E_1/E_2 =  10.74$, while $E_0/E_2 =  1.61\times 10^{3}$. Thus, we need at least ten times fewer expected transmissions to reconstruct. In this case, $E_3/E_2 = 8.6$ (and the $E_i$ are increasing for $i \geq 3$), so on average, outputs of length $n+2$ will be used for reconstruction. More generally, for $\ell$ small compared to $n$, if $p(n-\ell+1) > \ell \times \binom{2\ell}{\ell}$, $E_{\ell}$ is smaller than $E_{i}$ for $i < \ell$, and reconstruction will be performed prior to decoding.
}

\section{Conclusion}
In this work, we examined the exact reconstruction of sequences that are codewords of synchronization (insertion/deletion-correcting) codes from traces that are the result of an insertion channel. We provided exact formulas for the number of traces necessary for the binary and non-binary cases of this problem. These formulas resolve a problem left open by Levenshtein, who derived the first expressions for the uncoded case. We also examined traces produced by other channels, such as the insertion and deletion channel, {\color{black} and explored the application of the results to a probabilistic insertion channel}.

The expressions we found represent the worst-case number of traces needed when performing reconstruction in any code with the required minimum edit distance. We asked whether selecting a particular code allows us to reconstruct with fewer traces compared to the worst-case. We showed that for the popular single insertion/deletion-correcting Varshamov-Tenengolts codes, there are always many codeword pairs that require the worst-case number of traces for reconstruction. This inspires us to ask whether we can construct new codes that have similar properties to the VT codes, but better (smaller) requirements for reconstruction.

Our results can be viewed as a promising first step towards a more general theory for coded data reconstruction. This is a rich area with many interesting further questions. {\color{black} It would be interesting to allow the traces to be affected by other types of errors, such as deletions, substitutions, burst errors, transpositions, and others. It would be especially interesting to derive the maximal number of traces for the cases of deletion channels (for $\ell > 2$) and combined insertions/deletions/substitutions channels, which accurately model real-life data reconstruction scenarios.} Equally intriguing is a study of efficient algorithms for reconstruction given the necessary number of traces: for example, given $N^+_q(n,t,k,\ell)+1$ traces of $X$, what is the most efficient algorithm to reproduce $X$?

\section{Appendix}

\subsection{Proof of Lemma ~\ref{lem:recGen}}
We present a proof of Lemma~\ref{lem:recGen}, which we restate below: \\ \\
%\begin{lemma}
{\bf Lemma 6.}
%\label{lem:recGenApp}
\textit{
For $n\geq 1, q \geq 2$ and $t,k,\ell \geq 1$ with $t \geq k \geq \ell$, $\cN^+_q(n,t,k,\ell)$ satisfies the recursions
\begin{align*}\cN^+_q&(n,t,k,\ell) = \\
& \cN^+_q(n-1,t,k,\ell) + (q-1)\cN^+_q(n,t-1,k-1,\ell),\end{align*}
and
\begin{align*}&\cN^+_q(n,t,k,\ell) = \cN^+_q(n,t-1,k,\ell) + \\
&\enskip \cN^+_q(n-1,t,k-1,\ell-1) + (q-2) \cN^+_q(n,t-1, k-1, \ell) .\end{align*}
}
%\end{lemma}
\begin{IEEEproof}
The proofs of these formulas only use standard sum manipulations and binomial identities. We first show the series of equalities and then describe the steps. For the first recursion, we have
\begin{align*}
\cN^+_q&(n-1,t,k,\ell) + (q-1)\cN^+_q(n,t-1,k-1,\ell) \\
%%%%%%%%%%%%%%%%%%%%%%%%%%%%%%%%%%%%%
&\stackrel{(a)}{=}  \sum_{j=\ell}^k \sum_{i=0}^{k-j} \binom{t-k+2j}{j}\binom{t+j-i}{t-k+2j} \times \\
& \pushright{2  \binom{(n-1)+t}{i} (q-1)^i (-1)^{k+j-i} + (q-1) \times} \\
& \pushright{\left[  \sum_{j=\ell}^{k-1} \sum_{i=0}^{k-1-j} \binom{(t-1)-(k-1)+2j}{j} \times \right.} \\
& \left. \binom{(t-1)+j-i}{t-k+2j} \binom{n+t-1}{i} (q-1)^i (-1)^{k-1+j-i} \right]\\
%%%%%%%%%%%%%%%%%%%%%%%%%%%%%%%%%%%%%
&\stackrel{(b)}{=}  \sum_{j=\ell}^k \sum_{i=0}^{k-j} \binom{t-k+2j}{j}\binom{t+j-i}{t-k+2j} \times \\
& \pushright{\binom{n-1+t}{i} (q-1)^i (-1)^{k+j-i}} \\
&\enskip +  \sum_{j=\ell}^{k-1} \sum_{i=0}^{k-1-j} \binom{t-k+2j}{j}\binom{t-1+j-i}{t-k+2j} \times \\
& \pushright{\binom{n+t-1}{i} (q-1)^{i+1} (-1)^{k-1+j-i}} \\
%%%%%%%%%%%%%%%%%%%%%%%%%%%%%%%%%%%%%
&\stackrel{(c)}{=}  \sum_{j=\ell}^k \sum_{i=0}^{k-j} \binom{t-k+2j}{j}\binom{t+j-i}{t-k+2j} \times \\
&\pushright{ \binom{n-1+t}{i} (q-1)^i (-1)^{k+j-i}} \\
&\enskip   +  \sum_{j=\ell}^{k-1} \sum_{i=1}^{k-j} \binom{t-k+2j}{j}\binom{t+j-i}{t-k+2j} \times \\
&\pushright{ \binom{n+t-1}{i-1} (q-1)^{i} (-1)^{k+j-i}} \\
%%%%%%%%%%%%%%%%%%%%%%%%%%%%%%%%%%%%%
&\stackrel{(d)}{=}  \sum_{j=\ell}^k \sum_{i=1}^{k-j} \binom{t-k+2j}{j}\binom{t+j-i}{t-k+2j} \times \\
&\pushright{\binom{n-1+t}{i} (q-1)^i (-1)^{k+j-i}} \\
&\enskip + \sum_{j=\ell}^k  \binom{t-k+2j}{j}\binom{t+j}{t-k+2j} (-1)^{k+j} \\
&\enskip   +  \sum_{j=\ell}^{k-1} \sum_{i=1}^{k-j} \binom{t-k+2j}{j}\binom{t+j-i}{t-k+2j} \times \\
&\pushright{\binom{n+t-1}{i-1} (q-1)^{i} (-1)^{k+j-i}} \\
%%%%%%%%%%%%%%%%%%%%%%%%%%%%%%%%%%%%%
&\stackrel{(e)}{=}  \sum_{j=\ell}^{k-1} \sum_{i=1}^{k-j} \binom{t-k+2j}{j}\binom{t+j-i}{t-k+2j} \times \\ 
&\pushright{ (q-1)^i (-1)^{k+j-i} \left[  \binom{n-1+t}{i} + \binom{n+t-1}{i-1} \right] }\\
&\enskip + \sum_{j=\ell}^{k} \binom{t-k+2j}{j}\binom{t+j}{t-k+2j} (-1)^{k+j}  \\ %+ \binom{t+k}{k}\binom{t+k}{t-k+2k} (-1)^{2k}  \\
%%%%%%%%%%%%%%%%%%%%%%%%%%%%%%%%%%%%%
&\stackrel{(f)}{=}  \sum_{j=\ell}^{k-1} \sum_{i=1}^{k-j} \binom{t-k+2j}{j}\binom{t+j-i}{t-k+2j}\times \\
&\pushright{\binom{n+t}{i}  (q-1)^i (-1)^{k+j-i}}  \\
&\enskip+ \sum_{j=\ell}^{k} \binom{t-k+2j}{j}\binom{t+j}{t-k+2j} (-1)^{k+j}  \\
%&\quad + \sum_{j=\ell}^{k-1} \binom{t-k+2j}{j}\binom{t+j}{t-k+2j} (-1)^{k+j}  + \binom{t+k}{k}  \\
%%%%%%%%%%%%%%%%%%%%%%%%%%%%%%%%%%%%%
&\stackrel{(g)}{=}   \sum_{j=\ell}^k \sum_{i=0}^{k-j} \binom{t-k+2j}{j}\binom{t+j-i}{t-k+2j} \times \\
&\pushright{ \binom{n+t}{i} (q-1)^i (-1)^{k+j-i}\qquad} \\
&= \cN_q^+(n,t,k,\ell).
\end{align*} 
Step $(a)$ follows from the definition of $\cN$. In step $(c)$, we changed the range of summation for $i$ from $[0,k-j-1]$ to $[1,k-j]$ for the second term. In $(d)$, we broke up the sum for the first term, removing the components with $i=0$ in the inner sum. In step $(e)$, we note that there is no inner sum for $j=k$, so we change the limit of the outer sum to $k-1$. We then combined terms. In step $(f)$ we applied the identity $\binom{n+t-1}{i} + \binom{n+t-1}{i-1} = \binom{n+t}{i}$. All other steps are immediate rearrangements of terms.

Next, for the second recursion, we have that
\begin{align*}
&\cN^+_q(n,t-1,k,\ell) + \cN^+_q(n-1,t,k-1,\ell-1)  \\
& \enskip +(q-2) \cN^+_q(n,t-1, k-1, \ell) \\
&\stackrel{(a)}{=} \sum_{j=\ell}^k \sum_{i=0}^{k-j} \binom{(t-1)-k+2j}{j}\binom{(t-1)+j-i}{(t-1)-k+2j} \times \\
& \pushright{\binom{n+(t-1)}{i} (q-1)^i (-1)^{k+j-i}} \\
& \enskip + \sum_{j=\ell-1}^{k-1} \sum_{i=0}^{k-1-j} \binom{t-k+1+2j}{j}\binom{t+j-i}{t-(k-1)+2j} \times \\
& \pushright{\binom{(n-1)+t}{i} (q-1)^i (-1)^{k-1+j-i}} \\
& \enskip+ (q-2) \left[ \sum_{j=\ell}^{k-1} \sum_{i=0}^{k-1-j} \binom{t-1-(k-1)+2j}{j} \times \right. \\
& \left. \binom{t-1+j-i}{t-k+2j}\binom{n+t-1}{i} (q-1)^i (-1)^{k-1+j-i} \right]\\
%%%%%%%%%%%%%%%%%%%%%%%%%%%%%%%%%%%%%
%%%%%%%%%%%%%%%%%%%%%%%%%%%%%%%%%%%%%
&\stackrel{(b)}{=} \sum_{j=\ell}^k \sum_{i=0}^{k-j} \binom{t-k+2j-1}{j}\binom{t+j-i-1}{t-k+2j-1} \times \\
& \pushright{\binom{n+t-1}{i} (q-1)^i (-1)^{k+j-i}} \\
& \enskip+ \sum_{j=\ell}^{k} \sum_{i=0}^{k-j} \binom{t-k+2j-1}{j-1}\binom{t+j-i-1}{t-k+2j-1} \times  \\
&\pushright{\binom{n+t-1}{i} (q-1)^i (-1)^{k+j-i}} \\
&\enskip+ (q-2)  \left[  \sum_{j=\ell}^{k-1} \sum_{i=0}^{k-j-1} \binom{t-k+2j}{j}\binom{t+j-i-1}{t-k+2j} \times \right. \\
& \pushright{\left. \binom{n+t-1}{i} (q-1)^i (-1)^{k+j-i-1} \right]}\\
%%%%%%%%%%%%%%%%%%%%%%%%%%%%%%%%%%%%%
%%%%%%%%%%%%%%%%%%%%%%%%%%%%%%%%%%%%%
&\stackrel{(c)}{=}  \sum_{j=\ell}^k \sum_{i=0}^{k-j} \left[\binom{t-k+2j-1}{j} +  \binom{t-k+2j-1}{j-1} \right] \times \\
&\pushright{\binom{t+j-i-1}{t-k+2j-1} \binom{n+t-1}{i} (q-1)^i (-1)^{k+j-i}} \\
&\enskip+ (q-2)  \left[  \sum_{j=\ell}^{k-1} \sum_{i=0}^{k-j-1} \binom{t-k+2j}{j}\binom{t+j-i-1}{t-k+2j} \times \right. \\
&\pushright{\left. \binom{n+t-1}{i} (q-1)^i (-1)^{k+j-i-1} \right]}\\
%%%%%%%%%%%%%%%%%%%%%%%%%%%%%%%%%%%%%
&\stackrel{(d)}{=}  \sum_{j=\ell}^k \sum_{i=0}^{k-j} \binom{t-k+2j}{j} \binom{t+j-i-1}{t-k+2j-1} \times \\
&\pushright{ \binom{n+t-1}{i} (q-1)^i (-1)^{k+j-i}} \\
&\enskip+ (q-2) \left[  \sum_{j=\ell}^{k-1} \sum_{i=0}^{k-j-1} \binom{t-k+2j}{j}\binom{t+j-i-1}{t-k+2j} \times \right. \\
&\pushright{\left. \binom{n+t-1}{i} (q-1)^i (-1)^{k+j-i-1} \right]}\\
%%%%%%%%%%%%%%%%%%%%%%%%%%%%%%%%%%%%%
&\stackrel{(e)}{=}  \sum_{j=\ell}^k \sum_{i=0}^{k-j} \binom{t-k+2j}{j} \binom{t+j-i-1}{t-k+2j-1} \times \\
& \pushright{\binom{n+t-1}{i} (q-1)^i (-1)^{k+j-i}} \\
&\enskip -  \sum_{j=\ell}^{k-1} \sum_{i=0}^{k-j-1} \binom{t-k+2j}{j}\binom{t+j-i-1}{t-k+2j} \times \\
&\pushright{\binom{n+t-1}{i} (q-1)^{i} (-1)^{k+j-i-1}} \\
&\enskip+  \sum_{j=\ell}^{k-1} \sum_{i=0}^{k-j-1} \binom{t-k+2j}{j}\binom{t+j-i-1}{t-k+2j} \times \\
& \pushright{\binom{n+t-1}{i} (q-1)^{i+1} (-1)^{k+j-i-1}} \\
%%%%%%%%%%%%%%%%%%%%%%%%%%%%%%%%%%%%%%%%%%%%%%%%
&\stackrel{(f)}{=}  \sum_{j=\ell}^{k-1} \sum_{i=0}^{k-j-1} \binom{t-k+2j}{j} \times \\
& \pushright{\left[ \binom{t+j-i-1}{t-k+2j-1} + \binom{t+j-i-1}{t-k+2j}\right] \times }\\
&\pushright{ \binom{n+t-1}{i} (q-1)^i (-1)^{k+j-i} } \\
&\enskip+ \sum_{j=\ell}^k \binom{t-k+2j}{j} \binom{t+j-(k-j)-1}{t-k+2j-1}\times \\
& \pushright{\binom{n+t-1}{k-j} (q-1)^{k-j} (-1)^{k+j-(k-j)}} \\
&\enskip+  \sum_{j=\ell}^{k-1} \sum_{i=0}^{k-j-1} \binom{t-k+2j}{j}\binom{t+j-i-1}{t-k+2j}\times \\
& \pushright{\binom{n+t-1}{i} (q-1)^{i+1} (-1)^{k+j-i-1}} \\
%%%%%%%%%%%%%%%%%%%%%%%%%%%%%%%%%%%%%%%%%%%%%%%%
&\stackrel{(g)}{=}  \sum_{j=\ell}^{k-1} \sum_{i=0}^{k-j-1} \binom{t-k+2j}{j} \binom{t+j-i}{t-k+2j} \times \\
&\pushright{\binom{n+t-1}{i} (q-1)^i (-1)^{k+j-i}} \\
&\enskip+ \sum_{j=\ell}^k \binom{t-k+2j}{j} \binom{n+t-1}{k-j} (q-1)^{k-j} \\
&\enskip+  \sum_{j=\ell}^{k-1} \sum_{i=0}^{k-j-1} \binom{t-k+2j}{j}\binom{t+j-i-1}{t-k+2j} \times \\
&\pushright{ \binom{n+t-1}{i} (q-1)^{i+1} (-1)^{k+j-i-1}} \\
%%%%%%%%%%%%%%%%%%%%%%%%%%%%%%%%%%%%%%%%%%%%%%%%
&\stackrel{(h)}{=}  \sum_{j=\ell}^{k-1} \sum_{i=0}^{k-j-1} \binom{t-k+2j}{j} \binom{t+j-i}{t-k+2j} \times \\
&\pushright{\binom{n+t-1}{i} (q-1)^i (-1)^{k+j-i}} \\
& \enskip+ \sum_{j=\ell}^k \binom{t-k+2j}{j} \binom{n+t-1}{k-j} (q-1)^{k-j} \\
&\enskip +  \sum_{j=\ell}^{k-1} \sum_{i=1}^{k-j} \binom{t-k+2j}{j}\binom{t+j-i}{t-k+2j} \times \\
&\pushright{ \binom{n+t-1}{i-1} (q-1)^{i} (-1)^{k+j-i}} \\
%%%%%%%%%%%%%%%%%%%%%%%%%%%%%%%%%%%%%%%%%%%%%%%%
&\stackrel{(j)}{=}  \sum_{j=\ell}^{k-1} \sum_{i=1}^{k-j-1} \binom{t-k+2j}{j} \binom{t+j-i}{t-k+2j} \times \\
& \pushright{\left[ \binom{n+t-1}{i} +\binom{n+t-1}{i-1}\right] (q-1)^i (-1)^{k+j-i} }\\
&\enskip + \sum_{j=\ell}^{k-1} \binom{t-k+2j}{j} \binom{t+j}{t-k+2j} (-1)^{k+j} \\
&\enskip+ \sum_{j=\ell}^{k-1} \binom{t-k+2j}{j} \binom{n+t-1}{k-j-1} (q-1)^{k-j} \\
& \enskip+ \sum_{j=\ell}^k \binom{t-k+2j}{j} \binom{n+t-1}{k-j} (q-1)^{k-j} \\
%%%%%%%%%%%%%%%%%%%%%%%%%%%%%%%%%%%%%%%%%%%%%%%%
&\stackrel{(k)}{=}  \sum_{j=\ell}^{k-1} \sum_{i=1}^{k-j-1} \binom{t-k+2j}{j} \binom{t+j-i}{t-k+2j}  \times \\
&\pushright{\binom{n+t}{i}  (q-1)^i (-1)^{k+j-i}} \\
& \enskip+ \sum_{j=\ell}^{k-1} \binom{t-k+2j}{j} \binom{t+j}{t-k+2j} (-1)^{k+j} \\
&\enskip+ \sum_{j=\ell}^{k} \binom{t-k+2j}{j} \binom{n+t}{k-j} (q-1)^{k-j} \\
%%%%%%%%%%%%%%%%%%%%%%%%%%%%%%%%%%%%%%%%%%%%%%%%
&\stackrel{(l)}{=}  \sum_{j=\ell}^{k} \sum_{i=0}^{k-j} \binom{t-k+2j}{j} \binom{t+j-i}{t-k+2j}  \times \\
&\pushright{\binom{n+t}{i} (q-1)^i (-1)^{k+j-i}} \\
&= \cN_q^+(n,t,k,\ell).
\end{align*}

The steps we used are the following. In $(a)$ we applied the definition of $\cN$. In $(b)$, we changed the range of summation for $j$ in the middle term from $[\ell-1,k-1]$ to $[\ell,k]$. In $(d)$, we used the identity $\binom{t-k+2j-1}{j}+\binom{t-k+2j-1}{j-1} = \binom{t-k+2j}{j}$. In $(e)$, we broke up the second term from $(d)$, which is multiplied by a factor of $(q-2)$ into two terms, one multiplied by a factor of $(q-1)$ and the other by $(-1)$. In $(f)$, we combined the first two terms from $(e)$. In $(g)$, we used the identity $\binom{t+j-i-1}{t-k+2j-1} + \binom{t+j-i-1}{t-k+2j} = \binom{t+j-i}{t-k+2j}$. In $(h)$, we changed the range of summation for $i$ in the second term from $[0,k-j-1]$ to $[1,k-j]$. In $(j)$, we combined terms and again applied the identity $\binom{n+t-1}{i} + \binom{n+t-1}{i-1} = \binom{n+t}{i}$. We also combined the last two summands, using the identity $\binom{n+t-1}{k-j-1} + \binom{n+t-1}{k-j} = \binom{n+t}{k-j}$. In $(k)$ we combined all remaining terms. 
\end{IEEEproof}

\subsection{Proof of Lemma~\ref{lem:aux}}
We present the proof of the two auxiliary combinatorial identities. \\ \\
\textbf{Lemma 8.}
%\begin{lemma} \label{lem:aux} 
\textit{
   \begin{enumerate}[label={\arabic*.},wide, labelwidth=!, labelindent=0pt]
       \item For $m\geq 0$, 
		\[\sum_{j=0}^m \binom{2j}{j} \binom{m+j}{2j} (-1)^{m+j} = 1.\] \label{partone}
       \item For $n, m, t, j \geq 0$ and $t+j \geq m$,  
        \[\sum_{i=0}^m \binom{t+j-i}{t+j-m} \binom{n+t}{i} (-1)^{m-i} = \binom{n+m-j-1}{m}.\]
        \label{parttwo}
   \end{enumerate}
   }
%\end{lemma}
\begin{IEEEproof}
Both identities will be proved by a generating function approach. This strategy is described as the ``snake oil method'' in \cite{Wilf}. The idea is that the right-hand side of each identity has an easily-derived generating function, while we will perform more complex manipulations to derive an identical generating function for the left-hand side. For the first identity, the generating function $F(x)$ for the left-hand side is written as
\begin{align*}
F(x) &= \sum_{m \geq 0} x^m \sum_{j=0}^m \binom{2j}{j} \binom{m+j}{2j} (-1)^{m+j} \\
&= \sum_{j=0}^\infty \sum_{m \geq j} x^m  \binom{2j}{j} \binom{m+j}{2j} (-1)^{m+j} \\
&= \sum_{j = 0}^\infty \binom{2j}{j} x^{-j} \sum_{m \geq j} \binom{m+j}{2j} (-x)^{m+j} \\
&= \sum_{j = 0}^\infty \binom{2j}{j} x^{-j} \sum_{r' \geq 0} \binom{r'}{2j} (-x)^{r'} \\
&= \sum_{j = 0}^\infty \binom{2j}{j} x^{-j} \frac{(-x)^{2j}}{(1+x)^{2j+1}} \\
&= \frac{1}{1+x} \sum_{j = 0}^\infty \binom{2j}{j} \left( \frac{x}{(1+x)^2}\right)^{j} \\
&= \frac{1}{1+x} \frac{1}{\sqrt{1-\frac{4x}{(1+x)^2}}} \\
&= \frac{1}{1+x} \frac{1+x}{1-x} = \frac{1}{1-x}.
\end{align*}
In the fourth step, we replace $m+j$ with $r'$. We can start the sum at $r' = 0$ since the binomial term $\binom{r'}{2j}$ evaluates to 0 for all $m < j$. Next, in the fifth step, we use the series $\sum_{r \geq 0} \binom{r}{k} x^r = \frac{x^k}{(1-x)^{k+1}}$ \cite{Wilf}. The only condition for this identity is $2j \geq 0$. Next, in the seventh step, we applied the generating function for the central binomial coefficients \cite{Wilf}: \[\sum_{j \geq 0} \binom{2j}{j} x^j = \frac{1}{\sqrt{1-4x}}.\] Thus we conclude that $F(x) = 1+x+x^2 + \ldots$, so indeed $\sum_{j=0}^m \binom{2j}{j} \binom{m+j}{2j} (-1)^{m+j} = 1$. 

We use the same approach for the second identity. The right-hand side of the identity counts the number of ways to distribute $m$ items in $n-j$ buckets. It is easy to see that this quantity has, with respect to $m$, the generating function $(1+x+x^2+\ldots)^{n-j} = (1-x)^{-(n-j)}$. The left-hand side has generating function
\begin{align*}
F(x) &= \sum_{m \geq 0} x^m \sum_{i=0}^m \binom{t+j-i}{t+j-m} \binom{n+t}{i} (-1)^{m-i} \\
&= \sum_{i=0}^\infty \binom{n+t}{i} \sum_{m \geq i} x^m  \binom{t+j-i}{t+j-m}  (-1)^{m-i} \\
&= \sum_{i=0}^\infty \binom{n+t}{i} x^{i} \sum_{r \geq 0} x^r \binom{t+j-i}{t+j-(r+i)} (-1)^r \\
&= \sum_{i=0}^\infty \binom{n+t}{i} x^{i} \sum_{r \geq 0} x^r \binom{t+j-i}{r} (-1)^r \\
&= \sum_{i=0}^\infty \binom{n+t}{i} x^{i} (1-x)^{t+j-i} \\
&= (1-x)^{t+j} \sum_{i=0}^\infty \binom{n+t}{i} \left(\frac{x}{1-x}\right)^i \\
&= (1-x)^{t+j} \left( 1 + \frac{x}{1-x} \right)^{n+t} \\
&= (1-x)^{t+j} (1-x)^{-(n+t)} \\
&= (1-x)^{j-n}, 
\end{align*}
and we are done. In the third step, we write $m-i=r$. In the fifth and seventh steps, we applied the binomial theorem.
\end{IEEEproof}

{\color{black}
\subsection{Proof of Lemma~\ref{lem:Fprop}}
\textbf{Lemma 13.}
%\begin{lemma} \label{lem:aux} 
For any integer $n \geq 7$, and integer $0 \leq m \leq n$, there exists a subset $T' \subseteq FP(n)$ where $\sum_{i \in T'} i = m $. In addition, $\lceil \log_2(n) \rceil \leq |FP(n)| \leq \lceil \log_2(n) \rceil + 1$. Furthermore, if $n$ is even, we have $\{ \frac{n}{2}, \frac{n}{2} + 1\} \not \in FP(n)$ and if $n$ is odd, then $\{ \lfloor \frac{n}{2} \rfloor,\lfloor \frac{n}{2} \rfloor+2 \} \not \in FP(n)$. 
	
	\begin{IEEEproof} The third sentence in Lemma~\ref{lem:Fprop} is trivially true based on the steps in the algorithm. There are four possible cases of $ FP(n) $, dependent on $ n $. These cases follow directly from step 2 of the algorithm. We will show that each of these cases satisfies the conditions of Lemma~\ref{lem:Fprop}. In particular, for each case we will show that there exists a subset $T' \subseteq FP(n)$ where $\sum_{i \in T'} i = m $ (for $0 \leq m\leq n $), and that $\lceil \log_2(n) \rceil \leq |FP(n)| \leq \lceil \log_2(n) \rceil + 1$. In the following analysis, we make extensive use of the fact that there exists a subset of the set $ \{2^0,2^1,\ldots,2^k\} $ that sums to every positive integer up to $ 2^{k+1}-1. $ For the following cases, assume $ n\geq7 $ and $k$ is an integer.
		
	\textit{Case 1:} $ n=2^k $ or $ n=2^k+1 $. In this case, we have $ FP(n)=\{2^0,2^1,\ldots,2^{k-2},2^{k-1}-1,2^k-1\} $. We have that $ |FP(n)|=k+1 $, which satisfies the size condition in Lemma~\ref{lem:Fprop}. Note that the first $ k-1 $ elements can sum to any integer up to $ 2^{k-1}-1 $. Including the next element in $ FP(n) $, $ 2^{k-1}-1 $, we can now sum up to any integer $ 2(2^{k-1}-1)=2^k-2 $. Including the last element in the set, we see that there are no gaps and there exists a subset of $ FP(n) $ that sums up to any integer $0 \leq m\leq n $.
	
	\textit{Case 2:} $ n=2^k-2 $. In this case, we have $ FP(n)=\{2^0,2^1,\ldots,2^{k-2},2^{k-1}-2,2^k-2\} $. We have that $ |FP(n)|=k+1=\lceil \log_2(n) \rceil + 1 $, which satisfies the size condition in Lemma~\ref{lem:Fprop}. Again, note that the first $ k-1 $ elements can sum to any integer up to $ 2^{k-1}-1 $. Including the next element in $ FP(n) $, $ 2^{k-1}-2 $, we can now sum up to any integer $ 2^{k-1}-1+2^{k-1}-2=2^k-3 $. The final element in $ FP(n)=2^k-2 $, which is equal to $ n $ itself.
	
	\textit{Case 3:} $ n=2^k-3 $. In this case, we have $ FP(n)=\{2^0,2^1,\ldots,2^{k-2},2^{k-1}-1\} $. We have that $ |FP(n)|=k=\lceil \log_2(n) \rceil $, which satisfies the size condition in Lemma~\ref{lem:Fprop}. Again, note that the first $ k-1 $ elements can sum to any integer up to $ 2^{k-1}-1 $. Including the next element in $ FP(n) $, $ 2^{k-1}-1 $, we can now sum up to any integer $ 2(2^{k-1}-1)=2^k-2 $. The final element in $ FP(n)=2^k-3 $, which is equal to $ n $ itself.
	
	\textit{Case 4:} All other $ n\geq 7 $. In this case, we simply have $ FP(n)=\{2^0,2^1,\ldots,2^{\floor{\log(n)}}\} $. We have that $ |FP(n)|=k+1=\lceil \log_2(n) \rceil $. Additionally, since this set is simply all the powers of 2 less than $ n $, we know that there exists a subset of $ FP(n) $ that can sum to any integer $0 \leq m\leq n $. \end{IEEEproof}
	}
	
\begin{IEEEbiographynophoto}{Frederic Sala}
(S'13) received the B.S.E. degree in Electrical Engineering from the University of Michigan, Ann Arbor, in 2010 and the M.S. degree in Electrical Enginering from the University of California, Los Angeles (UCLA) in 2013. He is currently pursuing the Ph.D. degree in Electrical Engineering at UCLA, where he is associated with the LORIS and CoDESS labs.

His research interests include information theory, coding theory, and algorithms, including applications to data management, data synchronization, and learning. He is a recipient of the NSF Graduate Research Fellowship and the UCLA Edward K. Rice Outstanding Masters Student Award.
\end{IEEEbiographynophoto}

\begin{IEEEbiographynophoto}{Ryan Gabrys}
is a scientist at Spawar Systems Center Pacific. In 2010 he received the Master of Engineering degree from the University of California at San Diego and in 2014 he obtained a Ph.D. from the University of California at Los Angeles. His research interests include coding theory and its applications to storage and synchronization."
\end{IEEEbiographynophoto}
%\vfill

\begin{IEEEbiographynophoto}{Clayton Schoeny}
(S'09) is a Ph.D. student in the Electrical Engineering Department at the University of California, Los Angeles (UCLA). He received his B.S. and M.S. degrees in Electrical Engineering from UCLA in 2012 and 2014, respectively. He has industry experience with The Aerospace Corporation, DIRECTV, and SPAWAR.

His research interests include coding theory and information theory, and he is associated with the LORIS and CoDESS labs. He is a recipient of the Henry Samueli Excellence in Teaching Award.
\end{IEEEbiographynophoto}

\begin{IEEEbiographynophoto}{Lara Dolecek}
(S'05--M'10--SM'12) is an Associate Professor  with the Electrical Engineering Department at the University of California, Los Angeles (UCLA). She holds a B.S. (with honors), M.S. and Ph.D. degrees in Electrical Engineering and Computer Sciences, as well as an M.A. degree in Statistics, all from the University of California, Berkeley. She received the 2007 David J. Sakrison Memorial Prize for the most outstanding doctoral research in the Department of Electrical Engineering and Computer Sciences at UC Berkeley. Prior to joining UCLA, she was a postdoctoral researcher with the Laboratory for Information and Decision Systems at the Massachusetts Institute of Technology. She received IBM Faculty Award (2014), Northrop Grumman Excellence in Teaching Award (2013), Intel Early Career Faculty Award (2013), University of California Faculty Development Award (2013), Okawa Research Grant (2013), NSF CAREER Award (2012), and Hellman Fellowship Award (2011). With her research group, she received the best paper award from IEEE Globecom 2015 conference. Her research interests span coding and information theory, graphical models, statistical algorithms, and computational methods, with applications to emerging systems for data storage, processing, and communication. She currently serves as an Associate Editor for IEEE Transactions on Communications.	
\end{IEEEbiographynophoto}

\vfill

\end{document}